\documentclass[11pt,a4paper]{article}
\pdfoutput=1
\usepackage[utf8]{inputenc}
\usepackage{jheppub}
\usepackage{enumitem}
\usepackage{shuffle}
\usepackage{xcolor} 
\definecolor{green}{rgb}{0,0.69,0.2} 
\allowdisplaybreaks

\def\be{\begin{equation}}
\def\ee{\end{equation}}
\def\bea{\begin{eqnarray}}
\def\eea{\end{eqnarray}}
\def\beal{\begin{equation}\begin{aligned}}
\def\eeal{\end{aligned}\end{equation}}
\def\nn{\nonumber}

\def\u#1{\underline{#1}}
\def\o#1{\overline{#1}}

\def\Tr{\operatorname*{Tr}}
\def\tr{\operatorname*{tr}}
\def\tf{\tilde{f}}
\def\Res_#1{\operatorname*{Res}_{#1}}

\def\ie{i.~e. }
\def\eg{e.~g. }

\def\eqn#1{eq.~\eqref{#1}}
\def\eqns#1#2{eqs.~\eqref{#1} and~\eqref{#2}}

\def\fig#1{figure~{\ref{#1}}}

\def\Fig#1{Figure~{\ref{#1}}}

\def\sec#1{Section~{\ref{#1}}}
\def\secs#1#2{Section~{\ref{#1}} and~{\ref{#2}}}
\def\app#1{Appendix~{\ref{#1}}}
\def\rcite#1{ref.~\cite{#1}}
\def\rcites#1{refs.~\cite{#1}}

\def\qo#1{{\color{#1}\{}}
\def\qc#1{{\color{#1}\}}}
\def\qob#1{{\bf \color{#1}[\;\!}}
\def\qcb#1{{\bf \;\!\color{#1}]}}

\newcommand{\scalegraph}[2]{\vcenter{\hbox{\!\;\includegraphics[scale=#1]{Graphics/#2.pdf}\!\;}}}

\title{Multi-Quark Colour Decompositions from Unitarity}

\author[a]{Alexander Ochirov}
\author[b]{and Ben Page}

\affiliation[a]{ETH Z\"urich, Institut f\"ur Theoretische Physik,
Wolfgang-Pauli-Str. 27, 8093 Z\"urich, Switzerland}
\affiliation[b]{Institut de Physique Th\'eorique, CEA, CNRS,
Universit\'e Paris-Saclay, \\ F-91191 Gif-sur-Yvette cedex, France}

\emailAdd{aochirov@phys.ethz.ch}
\emailAdd{bpage@ipht.fr}

\abstract{Any loop QCD amplitude at full colour is constructed from kinematic and gauge-group building blocks. In a unitarity-based on-shell framework, both objects can be reconstructed from their respective counterparts in tree-level amplitudes. This procedure is at its most powerful when aligned with flexible colour decompositions of tree-level QCD amplitudes. In this note we derive such decompositions for amplitudes with an arbitrary number of quarks and gluons from the same principle that is used to bootstrap kinematics --- unitarity factorisation. In the process we formulate new multi-quark bases and provide closed-form expressions for the new decompositions. We then elaborate upon their application in colour decompositions of loop multi-quark amplitudes.
}

\preprint{IPhT-19/090}

\begin{document}
\maketitle

\section{Introduction}
\label{sec:intro}

The improving precision of the experimental measurements of multi-jet processes
at the Large Hadron Collider
has motivated an array of new theoretical results
for QCD scattering amplitudes with more than two partons in the final state~\cite{Gehrmann:2015bfy,Dunbar:2016aux,Dunbar:2016cxp,Dunbar:2016gjb,Dunbar:2017nfy,Badger:2017jhb,Abreu:2017hqn,
Chawdhry:2018awn,Badger:2018gip,Abreu:2018jgq,
Abreu:2018zmy,Abreu:2019odu,Badger:2019djh}.
In particular, the first non-planar five-point two-loop amplitude
has been recently computed for pure Yang-Mills theory in \rcite{Badger:2019djh},
starting from the full-colour integrand of \rcite{Badger:2015lda}.
This remarkably simple result for the five-gluon amplitude with all helicities
chosen positive reflects the known structure at the integrand
level~\cite{Badger:2013gxa,Badger:2015lda,Badger:2016ozq}.

The demand for accuracy will inevitably require going beyond the leading-colour
approximation at two loops. Furthermore, many relevant two-loop observables, 
such as Higgs plus jets, have non-planar contributions even in this approximation,
and so a thorough understanding of the gauge-group degrees of freedom
is essential. 
On-shell methods, \ie based on unitarity cuts~\cite{Bern:1994zx,Bern:1994cg,Britto:2004nc},
are a promising approach to compute such higher-multiplicity two-loop amplitudes, and the loop-colour method of~\rcites{Badger:2015lda,Ochirov:2016ewn} provides a systematic on-shell method for handling colour information at the multi-loop level.
In a nutshell, the method consistently retains the colour factors of the tree
amplitudes of generalised unitarity cuts when reconstructing the loop
integrand.
It was detailed in \rcite{Ochirov:2016ewn} with a focus on the case of
purely adjoint-representation particle content.

A key ingredient for the loop-colour method presented in \rcite{Ochirov:2016ewn}
was a flexible tree-level colour decomposition.
In the adjoint case, the method relies on
the decomposition of del Duca, Dixon and Maltoni (DDM)~\cite{DelDuca:1999rs}.
This decomposition is a ``proper''
colour decomposition --- it splits the colour and kinematic degrees of
freedom in such a way that the kinematic objects are linearly independent planar
ordered amplitudes, in which only factorisation channels with consecutively
ordered particles may appear (see \eg \rcite{Dixon:1996wi}).
More precisely, the DDM decomposition expresses a purely gluonic amplitude
in terms of a basis of $(n-2)!$ ordered amplitudes independent under
Kleiss-Kuijf (KK) relations~\cite{Kleiss:1988ne} where the associated colour
factors are given by strings of the structure constants
\vspace{-18.5pt}
\beal
   {\cal A}^\text{tree}_n = \!\sum_{\sigma \in S_{n-2}}\!\!
   \begin{aligned} & \\
      \tf^{\,a_1 a_{\sigma(2)} b_1} \tf^{\,b_1 a_{\sigma(3)} b_2} \dots
      \tf^{\,b_{n-4} a_{\sigma(n-2)} b_{n-3}}
      \tf^{\,b_{n-3} a_{\sigma(n-1)} a_n} & \\ \times
      A(1,\sigma(2),\dots,\sigma(n-1),n) &
   \end{aligned} & \\
    = \!\sum_{\sigma \in S_{n-2}}\!\!
      C\Bigg(\scalegraph{0.75}{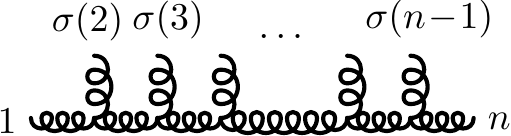}\Bigg)\,
      A(1,\sigma(2),\dots,\sigma(n-1),n) &.
\label{DDM}
\eeal
Note that in the ordered amplitudes the positions of gluons $1$ and $n$ are
fixed next to each other, and the topology of the corresponding colour factors
can be understood as ``stretched between'' these gluons. A similar colour decomposition
is known for tree amplitudes with an arbitrary number of quark
lines~\cite{Johansson:2015oia,Melia:2015ika} but restricted to the basis in
which two quarks of the same flavour are fixed next to each other
\cite{Melia:2013bta,Melia:2013epa}.

In this paper we give new tree-level colour decompositions for
multi-quark amplitudes, which allow any pair of particles to be fixed next to
each other in the ordered-amplitude sum. These decompositions are also
``proper'', sharing the crucial property of the DDM formula~\eqref{DDM}:
the constituent ordered amplitudes are independent with respect to the
KK relations.
Importantly, we find that once such bases of ordered amplitudes are found,
the colour factors in these decompositions are easily fixed
by imposing the principle of kinematic factorisation.
Remarkably,
the colour factors in such decompositions
obey factorisation relations themselves.
This then transparently explains the ``stretched'' nature
of the colour factors in the DDM decomposition~\eqref{DDM},
a feature inherited by all of the new decompositions in this work.
When inserted into unitarity cuts,
these results allow one to realise the full power
of the loop-colour method~\cite{Ochirov:2016ewn}
from pure Yang-Mills theory to QCD.

Note that we only consider the colour algebra defined with the most generic
set of relations --- the Jacobi and commutation relations ---
avoiding the details of the kinematic amplitude dependence.
Therefore, all our results are not specific to quarks but hold for other gauged matter particles in arbitrary representations of the gauge group,
whose flavour is conserved in gauge interactions.\footnote{In
other words, each flavoured colour-flow line may contain its own generators.
For example, what we refer to here
as the quark line $\u{1} \leftarrow \o{2}$ could just as well be
a complex scalar in the antifundamental representation of ${\rm U}(N_{\rm c})$,
meaning that its actual charge would flow against the arrow.}

The bulk of this paper is structured in a general-to-detailed way.
First, in \sec{sec:summary}, we describe the main result of this paper
--- KK-independent tree-level colour decompositions
for an arbitrary stretch, whose colour factors satisfy a set of
recursion relations.
Then in \sec{sec:tree}
we specify the main technical advance allowing these decompositions:
a collection of ``co-unitary'' KK-independent bases of tree-level amplitudes
that obey certain factorisation properties.
Next, in \sec{sec:loop}, we outline their application at loop level
within the method of \rcite{Ochirov:2016ewn}
and discuss the one-loop case~\cite{Bern:1990ux,Bern:1994fz,Ita:2011ar,
Reuschle:2013qna,Schuster:2013aya,Kalin:2017oqr}
in some detail.
Finally, we conclude by presenting our outlook in \sec{sec:outro}.

However, before we proceed in this way,
let us illustrate the appropriateness of a unitarity-based approach
to colour with a simple five-point example.

\subsection{Example of colour constraints from unitarity}
\label{sec:five}

In this section we use an amplitude with two quark pairs and one gluon
to show how a new colour decomposition can be constructed
from factorisation properties.
We label the first quark pair as $\u{1}$ and $\o{2}$,
the second quark pair as $\u{3}$ and $\o{4}$,
and the gluon as $5$.
The previously known colour decomposition~\cite{Johansson:2015oia}
would correspond to a $q\bar{q}$ stretch.
Here we wish to consider a $qg$ stretch.
Fixing for definiteness quark $\u{1}$ and gluon $5$ next to each other,
let us look for a tentative colour decomposition
\small
\begin{align}\label{A5qg}
   \scalegraph{0.86}{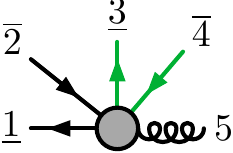} & \qquad \qquad \qquad\!\quad
   \scalegraph{0.86}{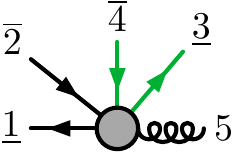} \qquad \qquad \qquad\!\quad
   \scalegraph{0.86}{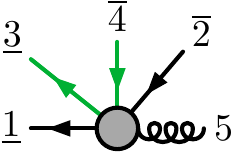} \!\!\!\!\\\!\!
   {\cal A}^{\text{tree}}_{5,2}\!
    = C(\u{1},\o{2},\u{3},\o{4},5) A(\u{1},\o{2},\u{3},\o{4},5) &
    + C(\u{1},\o{2},\o{4},\u{3},5) A(\u{1},\o{2},\o{4},\u{3},5)
    + C(\u{1},\u{3},\o{4},\o{2},5) A(\u{1},\u{3},\o{4},\o{2},5) . \nn
\end{align}
\normalsize
Here we give a graphic representation
of the cyclic amplitude orderings,
which will be used extensively throughout the paper.
For brevity, we have simply assumed that the orderings above constitute a valid
basis, although this can be argued from the presence
of the physical factorisation channels just as well.
For notational convenience in this paper we shall consider massless amplitudes,
though this is not required for the results to hold.

To constrain the unknown colour factors in \eqn{A5qg},
we consider the factorisation limits of the full amplitude
on the poles that include gluon~$5$ and one of the unfixed quarks:
\begin{subequations} \begin{align}
   \scalegraph{0.86}{AqqQQg1} & \xrightarrow[s_{45}\to0]{}
   \scalegraph{0.86}{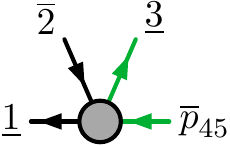} \times\frac{i}{s_{45}}\times \scalegraph{0.86}{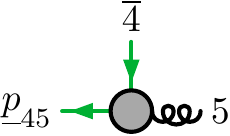} , \\
   \scalegraph{0.86}{AqqQQg2} & \xrightarrow[s_{35}\to0]{}
   \scalegraph{0.86}{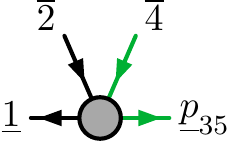} \times\frac{i}{s_{35}}\times \scalegraph{0.86}{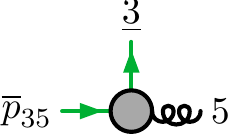} , \\
   \scalegraph{0.86}{AqqQQg3} & \xrightarrow[s_{25}\to0]{}
   \scalegraph{0.86}{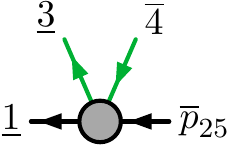} \times\frac{i}{s_{25}}\times \scalegraph{0.86}{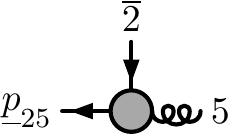} .
\end{align} \label{A5qgFactorization}%
\end{subequations}
As each pole only appears in one ordered amplitude,
the blob diagrams above can be regarded as either colour-dressed amplitudes
(in which case those on the left-hand side are all ${\cal A}^{\text{tree}}_{5,2}$)
or their ordered counterparts.
Now recall the trivial decomposition
of the four-point pure-quark amplitude\footnote{In our notation,
$C$ applied to a trivalent diagram (without blobs) translates to its colour factor
obtained using the colour Feynman rules recalled in \app{app:feynmanrules},
whereas $C$ applied to a blob diagram means the colour coefficient
of an ordered amplitude within the context of some colour decomposition.
These notions coincide in the factorisation limits that
expose the maximal number of propagators,
as the colour coefficients of three-point amplitudes
are always the same as the colour factors of the corresponding trivalent vertices.
}
\be
   {\cal A}^{\text{tree}}_{4,2}
    = C\!\left(\scalegraph{0.86}{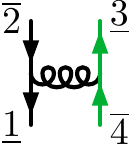}\right)
      A\;\!\!\!\left(\scalegraph{0.86}{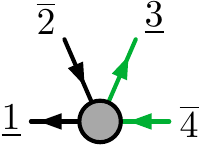}\right) ,
\label{A4qqQQ}
\ee
which must appear in the above factorisation limits of the five-point amplitude.
Applying the limits~\eqref{A5qgFactorization}
both to the five-point decomposition~\eqref{A5qg} as a whole
and to the constituent ordered amplitudes therein,
we can directly read off the colour coefficients as
\begin{subequations} \begin{align}
   C\!\left(\scalegraph{0.86}{AqqQQg1}\:\!\!\right) & =
      C\!\left(\scalegraph{0.86}{AqqQQ1}\right)
      C\!\left(\scalegraph{0.86}{AQQg1}\:\!\!\right)
    = C\!\left(\scalegraph{0.86}{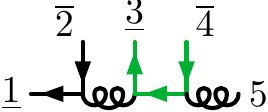}\right) , \\
   C\!\left(\scalegraph{0.86}{AqqQQg2}\:\!\!\right) & =
      C\!\left(\scalegraph{0.86}{AqqQQ2}\right)
      C\!\left(\scalegraph{0.86}{AQQg2}\:\!\!\right)
    = C\!\left(\scalegraph{0.86}{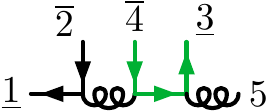}\right) , \\
   C\!\left(\scalegraph{0.86}{AqqQQg3}\:\!\!\right) & =
      C\!\left(\scalegraph{0.86}{AqqQQ3}\right)
      C\!\left(\scalegraph{0.86}{Aqqg25}\:\!\!\right)
    = C\!\left(\scalegraph{0.86}{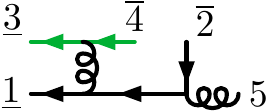}\right) .
\end{align} \label{A5qgColorFactorization}%
\end{subequations}
The kinematic limits here became identities,
since of course the colour factors are independent of any momenta.
In this way, we have used the factorisation properties of the amplitude
to derive a new explicit colour decomposition
\beal\!\!
   {\cal A}^{\text{tree}}_{5,2}
    = C\!\left(\scalegraph{0.86}{CqqQQg1}\right)
      A\;\!\!\!\left(\scalegraph{0.86}{AqqQQg1}\:\!\!\right)
    + C\!\left(\scalegraph{0.86}{CqqQQg2}\right)
      A\;\!\!\!\left(\scalegraph{0.86}{AqqQQg2}\:\!\!\right) & \\
   +\,C\!\left(\scalegraph{0.86}{CqqQQg3}\right)
      A\;\!\!\!\left(\scalegraph{0.86}{AqqQQg3}\:\!\!\right) & .
\label{A5qgFinal}
\eeal

\section{Colour and unitarity}
\label{sec:summary}

The basic principle that we shall be employing in this work is that
colour degrees of freedom play no kinematic role,
and so the operations of performing a colour decomposition
and taking a factorisation limit commute.
This implies strong constraints on any colour decomposition, and we will show
that, given a good choice of basis of ordered amplitudes, such
constraints can be used to fix it entirely.  We will prove the existence
of a collection of proper colour decompositions of the form
\begin{equation}
    \mathcal{A}(1,X,n) = \sum_{\sigma \in {\cal B}^{1,n}_X} 
      C\!\left(\scalegraph{0.86}{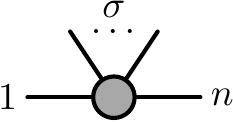}\right)
      A\!\left(\scalegraph{0.86}{A1n1permsExposed}\right) .
\end{equation}
Here we portray particles $1$ and $n$ as ``stretched'',
and we sum over a KK-independent basis,
indexed by the ordered tuples ${\cal B}^{1,n}_X$.
This quite abstract depiction is due to the generality of these
decompositions --- particles $1$ and $n$ can be taken arbitrarily.
That is, we will present decompositions for distinct-flavour
multi-parton tree-level amplitudes stretched across two gluons,
a gluon and a quark, and two quarks (which may be distinct).
This generalises the DDM decomposition~\cite{DelDuca:1999rs}
and that of \rcite{Johansson:2015oia} to a new level of flexibility.

As we shall see, the colour factors in these decompositions can be
entirely fixed by recursion relations obtained from factorisation
constraints.    Remarkably, we will show that the colour factors in our
bases obey a type of factorisation, being given as products of
colour factors of lower-point decompositions.
Specifically, if we consider a colour
factor in some decomposition where we have fixed particles $1$ and $n$
to be next to each other, then
\begin{equation}
   C\!\left(\scalegraph{0.86}{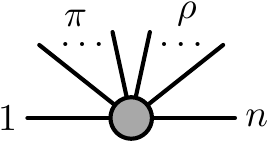}\right)
    = C\!\left(\scalegraph{0.86}{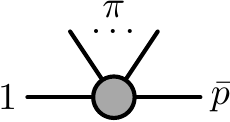}\right)
      C\!\left(\scalegraph{0.86}{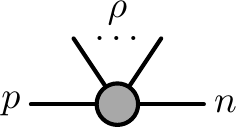}\right) .
\label{eq:ColourFactorization}
\end{equation}
That is, the colour factor of the ordered amplitude $A(1,\pi,\rho,n)$,
with $\pi$ and $\rho$ being ordered sets, factorises into the colour factors
of the ordered amplitudes $A(1,\pi,\bar{p})$ and $A(p,\rho,n)$,
where $p$ is the particle that allows the two factor amplitudes
to conserve all charges.
Importantly, $p$ may not exist, at which point
\eqn{eq:ColourFactorization} does not apply.
This may occur, for example,
where there is no way for a single particle to conserve quark flavour.

The colour factorisation relation~\eqref{eq:ColourFactorization} alone
is sufficient to fully fix the DDM decomposition.
Indeed, as $p$ is always a gluon in the purely gluonic case,
the colour factors in the DDM decomposition can be repeatedly split by
\eqn{eq:ColourFactorization} until one arrives at a product of three-point
colour factors --- the familiar ``comb'' structure of \eqn{DDM}.

In the presence of matter particles,
the abstract generality of the colour factorisation~\eqref{eq:ColourFactorization}
is a consequence of a crucial property of the bases presented in this work
(as well as the KK~\cite{Kleiss:1988ne} and Melia bases~\cite{Melia:2013bta,Melia:2013epa}),
which we dub ``co-unitarity''. This highly
non-trivial condition is that for every factorisation limit
that separates particles $1$ and $n$,
the surviving terms are independent under KK relations.
We define this notion more precisely, and write down a set of co-unitary bases in 
\sec{sec:tree}. Furthermore, we will find that any set of colour-ordered
amplitudes that satisfies this property is automatically a basis.

Beyond \eqn{eq:ColourFactorization}, we shall show further relations
between the factors of our decompositions, that completely fix the
colour factors in terms of three-point colour factors.
First, if there is a quark pair
with the quark and its antiquark next to each other,
then we find a ``leg-exchange'' relation,
\beal
   C\!\left(\scalegraph{0.86}{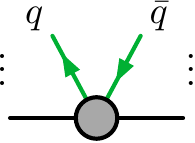}\right)
    - \theta \cdot C\!\left(\scalegraph{0.86}{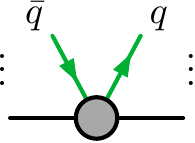}\right)
    = C\!\left(\scalegraph{0.86}{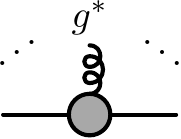}\right)
      C\!\left(\scalegraph{0.86}{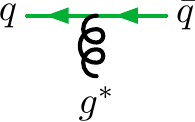}\right) .
\label{CqqFactorization2}
\eeal
Here $\theta$ reflects the fact that
the second term only contributes to the relation
if both associated ordered amplitudes appear in a colour decomposition.
As will be explained in detail in \sec{sec:tree}
and is familiar from the Melia basis~\cite{Melia:2013bta,Melia:2013epa},
a given colour decomposition may not include both orderings of a quark pair.
When only a single ordering is allowed, then $\theta=0$, otherwise $\theta=1$.
Similarly, we will show that
if we have a colour factor of an ordered amplitude
where the gluon is on one side of an (anti)quark,
we can move it to the other, at the price of generating an extra term:
\begin{subequations} \begin{align}
\label{CqgFactorization1}
   C\!\left(\scalegraph{0.86}{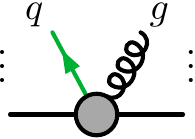}\right)
    - C\!\left(\scalegraph{0.86}{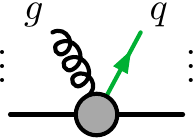}\right) &
    = C\!\left(\scalegraph{0.86}{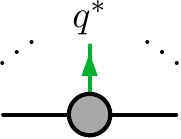}\right)
      C\!\left(\scalegraph{0.86}{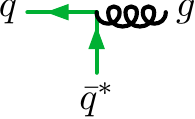}\right) , \\
\label{CqgFactorization2}
   C\!\left(\scalegraph{0.86}{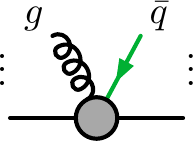}\right)
    - C\!\left(\scalegraph{0.86}{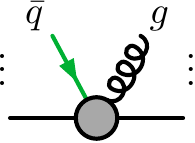}\right) &
    = C\!\left(\scalegraph{0.86}{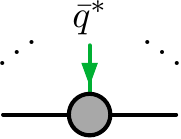}\right)
      C\!\left(\scalegraph{0.86}{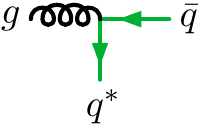}\right) .
\end{align} \label{CqgFactorization}%
\end{subequations}
Finally, the colour factors of two ordered amplitudes
that differ only by a permutation of two adjacent gluons
differ by a similar term:
\beal
   C\!\left(\scalegraph{0.86}{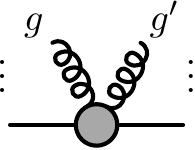}\right)
    - C\!\left(\scalegraph{0.86}{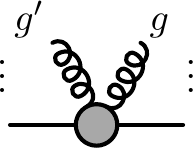}\right) &
    = C\!\left(\scalegraph{0.86}{AgExposed}\right)
      C\!\left(\scalegraph{0.86}{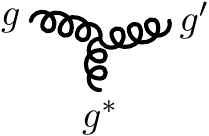}\right) .
\label{CggFactorization}
\eeal
Naturally,
these leg-exchange identities are a consequence
of the Jacobi identity and commutation relations,
\begin{subequations} \begin{align}
\label{Jacobi}
   \tf^{dac} \tf^{cbe} - \tf^{dbc} \tf^{cae} & = \tf^{abc} \tf^{dce} , \\
\label{Commutation}
   T^{a}_{i \bar \jmath} \, T^{b}_{j \bar k}   -
   T^{b}_{i \bar \jmath} \, T^{a}_{j \bar k} & = \tf^{abc} \, T^{c}_{i \bar k} .
\end{align} \label{JacobiCommutation}%
\end{subequations}
However, we shall see that they can also regarded as a consequence of factorisation.

These relations are a consequence of requiring that
the decompositions act trivially under various factorisation limits.
In the following subsections we shall discuss these limits
and the constraints that they put on the colour-ordered amplitudes in the bases.
We shall later see in \sec{sec:tree} that these constraints can indeed be realised.

\subsection{Colour factorisation}
\label{sec:factorization}

Let us now show that, given an appropriate basis of colour-ordered amplitudes,
the colour factorisation relation~\eqref{eq:ColourFactorization} indeed follows
from kinematic factorisation.
We will consider an $n$-particle amplitude, privileging particles $1$ and $n$ and
partitioning the remaining particles into the unordered sets $P$ and $R$. We now
consider a factorisation limit (implemented as a residue) on the channel
collecting all particles $1$ and $P$, and then colour-decompose the resulting
product:
\begin{align}
  \begin{split}
\label{eq:FactorizeThenDecompose}
 & \Res_{s_{1P}=0} {\cal A}\!\left(\scalegraph{0.86}{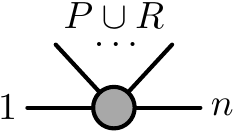}\right)
    = {\cal A}\!\left(\scalegraph{0.86}{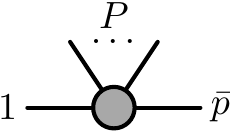}\right) \times
      {\cal A}\!\left(\scalegraph{0.86}{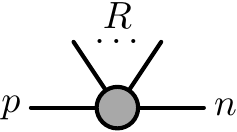}\right) \\ &~
   =\!\sum_{\pi \in {\cal B}_{P}^{1,\bar{p}}}
      \sum_{\rho \in {\cal B}_{R}^{p,n}}
      C\!\left(\scalegraph{0.86}{A1p1permExposed}\right)
      C\!\left(\scalegraph{0.86}{Apn1permExposed}\right)
      A\!\left(\scalegraph{0.86}{A1p1permExposed}\right)
      A\!\left(\scalegraph{0.86}{Apn1permExposed}\right) . 
  \end{split}
\end{align}
Here and below we use $\mathcal{A}$ to represent full-colour amplitudes,
$A$ to represent colour-ordered amplitudes,
and the notation ${\cal B}^{a,b}_{X}$ to indicate the permutation set
corresponding to the basis of ordered amplitudes
with particles $a$ and $b$ stretched across the set of particles $X$.
Importantly, there are no (kinematically independent) relations
between the ordered amplitudes on the right-hand side as the linear independence is inherited from the individual colour decompositions.
In order to find a constraint, we next consider the opposite order of operations ---
we first perform colour decomposition and then take the factorisation limit:
\beal
   \Res_{s_{1P}=0} {\cal A}\!\left(\scalegraph{0.86}{A1n2setsExposed}\right)
    = \Res_{s_{1P}=0} \sum_{\sigma \in {\cal B}_{P \cup R}^{1,n}}\!
      C\!\left(\scalegraph{0.86}{A1n1permsExposed}\right)
      A\!\left(\scalegraph{0.86}{A1n1permsExposed}\right) & \\
    =\!\!\!\sum_{(\sigma_1,\sigma_2) \in
            {\cal U}_{P,R}\!\big[{\cal B}_{P \cup R}^{1,n}\big]}\!\!
      C\!\left(\scalegraph{0.86}{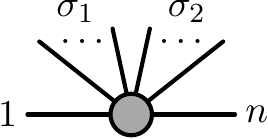}\right)
      A\!\left(\scalegraph{0.86}{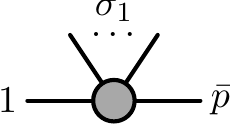}\right)
      A\!\left(\scalegraph{0.86}{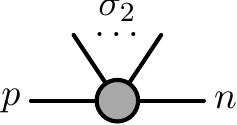}\right) & .
    \label{eq:DecomposeThenFactorize}
\eeal
Here the sum is over all terms in the colour decomposition
that have the chosen residue,
so the implied definition for the set
${\cal U}_{P,R}\big[{\cal B}_{P \cup R}^{1,n}\big]$ is
\be
   {\cal U}_{P,R}\big[{\cal B}_{P \cup R}^{1,n}\big]
    = \Big\{ (\pi,\rho) \in {\rm S}_P \times {\rm S}_R ~\Big| \quad
             \pi\oplus\rho \in {\cal B}_{P \cup R}^{1,n} , \quad
             \Res_{s_{1P}=0} A(1,\pi,\rho,n) \neq 0 \Big\} ,
\label{BasisFactorization}
\ee
where $\pi$ and $\rho$ are suborderings and
$\pi\oplus\rho$ is their concatenation.
In order to prove the colour-factorisation formula~\eqref{eq:ColourFactorization}
by equating \eqns{eq:FactorizeThenDecompose}{eq:DecomposeThenFactorize},
we need to be able to identify the corresponding
permutations $\sigma_1 = \pi$ and $\sigma_2 = \rho$.
For this to be true,
the two sets of cut permutations must be identical.
In other words, colour factorisation relies on an important property connecting
bases of different multiplicities
\be
   {\cal U}_{P,R}\big[{\cal B}_{P \cup R}^{1,n}\big]
    = {\cal B}_{P}^{1,\bar{p}} \times {\cal B}_{R}^{p,n} .
\label{CoUnitarity}
\ee
We dub a set of bases satisfying this constraint ``co-unitary.''
We construct such a set of bases for all multi-quark amplitudes in
\sec{sec:tree}.

From this derivation it is clear why the particle $p$ needs to exist:
otherwise the factorisation limits yield zero and therefore no constraint.
As previously mentioned, if $p$ were to always exist,
then we necessarily end up with a comb structure for colour factors,
which is not the case for the decomposition of \rcite{Johansson:2015oia}
corresponding to a $q\bar{q}$ stretch.
This occurs when the total flavour quantum numbers
of the set $\{1,P\}$ cannot be balanced by a single particle.
Nevertheless, we can still find constraints from kinematic factorisation.

\subsection{Leg-exchange relation and colour-ordered splitting}
\label{sec:colorunitarity2}

Let us now show that for the leg-exchange relations
\eqref{CqqFactorization2}--\eqref{CggFactorization} to hold,
we require a basis which respects ``colour-ordered splitting''.
We once again privilege two legs, $1$ and $n$, and now additionally pick
two other particles, $i$ and~$j$, in a colour-dressed amplitude.
On the one hand, if we take the corresponding two-particle factorisation limit
and then colour-decompose it, we obtain
\begin{align}
\label{FactorizeThenDecompose2}
 & \Res_{s_{ij}=0} {\cal A}\!\left(\scalegraph{0.86}{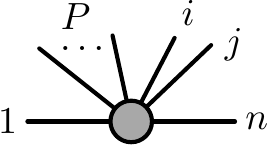}\right)
    = {\cal A}\!\left(\scalegraph{0.86}{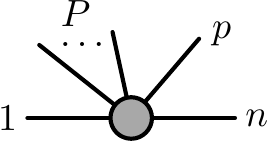}\right)
      {\cal A}\!\left(\scalegraph{0.86}{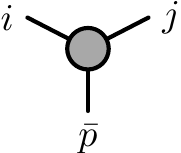}\right) \\* &~=
      \sum_{\substack{ (\pi_1\oplus(p)\oplus\pi_2) \\\;
                     =\,\pi \in {\cal B}_{P\cup\{p\}}^{1,n} }}
      C\!\left(\scalegraph{0.86}{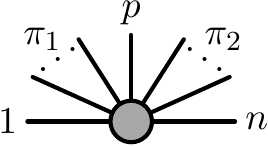}\right)
      C\!\left(\scalegraph{0.86}{A3ij}\right)
      A\!\left(\scalegraph{0.86}{Anp2permsAlt}\right)
      A\!\left(\scalegraph{0.86}{A3ij}\right) . \nn
\end{align}
Here the set of the unselected legs $P=\{2,\dots,n\!-\!1\}\setminus\{i,j\}$
is permuted along with the cut leg $p$,
and in the resulting permutations $\pi=\pi_1\oplus(p)\oplus\pi_2$
we track the position of $p$.

On the other hand, applying the same factorisation limit
to the already colour-decomposed amplitude yields
\beal
\label{DecomposeThenFactorize2}
   \Res_{s_{ij}=0} {\cal A}\!\left(\scalegraph{0.86}{AnPij}\right)
    = \Res_{s_{ij}=0} \sum_{\sigma \in {\cal B}_{P\cup\{i,j\}}^{1,n}}\!\!
      C\!\left(\scalegraph{0.86}{A1n1permsExposed}\right)
      A\!\left(\scalegraph{0.86}{A1n1permsExposed}\right) & \\
    = \sum_{\substack{ (\sigma_1\oplus(i,j)\oplus\sigma_2) \\\;
                     =\,\sigma \in {\cal B}_{P\cup\{i,j\}}^{1,n} }}
      C\!\left(\scalegraph{0.86}{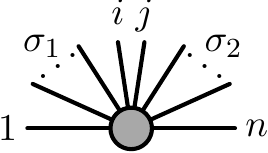}\right) \Res_{s_{ij}=0}
      A\!\left(\scalegraph{0.86}{Anij2perms}\right) & \\ +\!\!\!
      \sum_{\substack{ (\sigma_1\oplus(j,i)\oplus\sigma_2) \\*\;
                     =\,\sigma \in {\cal B}_{P\cup\{i,j\}}^{1,n} }}
      C\!\left(\scalegraph{0.86}{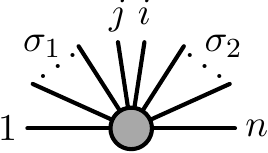}\right) \Res_{s_{ij}=0}
      A\!\left(\scalegraph{0.86}{Anji2perms}\right) &,
\eeal
where we consider only the colour-ordered amplitudes with particles~$i$ and~$j$
next to each other, as only they could possibly have a non-zero residue in that
limit. We note that a given colour decomposition may not actually include both
terms in \eqref{DecomposeThenFactorize2}.
Nevertheless, generically we can take this residue as
\be
   \Res_{s_{ij}=0} A\!\left(\scalegraph{0.8}{Anij2perms}\right)
    = -\!\!\Res_{s_{ij}=0} A\!\left(\scalegraph{0.8}{Anji2perms}\right)
    = A\!\left(\scalegraph{0.8}{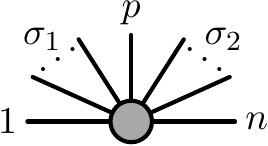}\right)
      A\!\left(\scalegraph{0.8}{A3ij}\right) .
\ee
Given this relation, we can rearrange equation \eqref{DecomposeThenFactorize2}
such that the kinematic part of the summand now takes the same form as that
of \eqref{FactorizeThenDecompose2}. In order to equate the colour factors of
these two representations, we need to be able to identify the set of
permutations. That is, it must be possible to construct all terms in the
$n$-point basis of colour-ordered amplitudes where $i$ and $j$ are found next to
each other by taking the basis of $(n-1)$-point amplitudes and replacing the
particle $p$ which balances all quantum numbers of the pair $(i,j)$. This
operation has a physical interpretation as an ordered ``splitting'' of the
particle~$p$ into the pair $(i,j)$, and so we require the bases to behave
naturally under splittings.
More precisely,
for any ordered pair $(i,j)$ among the unstretched particles in the $n$-point basis,
we require that the bases satisfy
\be
\label{eq:BasisSplittingConstraint}
   {\cal B}_{P\cup\{p\}}^{1,n} \big|_{p \to (i,j)} =
   \big\{ \sigma_1\oplus(i,j)\oplus\sigma_2
        = \sigma \in {\cal B}_{P\cup\{i,j\}}^{1,n}
   \big\} .
\ee
For a basis that satisfies this relation, the kinematic parts of
\eqns{FactorizeThenDecompose2}{DecomposeThenFactorize2} can be identified,
allowing us to equate the permutations $\sigma_i$ with $\pi_i$. In \sec{sec:ColourSplittingProof} we show that, up to the potential
vanishing of terms in this derivation, the bases of colour-ordered amplitudes
considered in this work indeed satisfy \eqref{eq:BasisSplittingConstraint},
leading us to the leg-exchange relations~\eqref{CqqFactorization2}--\eqref{CggFactorization}.

\section{Co-unitary tree-level bases and colour decompositions}
\label{sec:tree}

In this section we discuss various colour decompositions of tree-level QCD
amplitudes. Each decomposition corresponds to a basis of ordered amplitudes with
two chosen particles fixed next to each other.
These bases of amplitudes are independent with respect to the KK relations,
\ie over the field of rational numbers.
For tree-level amplitudes with $n$~particles and $k$~quark lines,
any such basis is composed from $(n-2)!/k!$ independent elements~\cite{Melia:2013epa}.
We leave discussion of amplitude bases with respect to
the kinematically dependent Bern-Carrasco-Johansson~\cite{Bern:2008qj,Johansson:2015oia} relations for future work.
Here we shall see that the colour factors associated with the KK-independent bases are given by the colour diagrams (or sets thereof) that resemble combs stretched by
the fixed particles from two sides.

\subsection{Like-flavour stretch}
\label{sec:qq}

For completeness and in order to introduce concepts that will be important at
later stages, we begin by reviewing the amplitude basis with two quarks of the
same flavour fixed next to each other introduced by
Melia in~\rcites{Melia:2013bta,Melia:2013epa}, as well as the corresponding colour
decomposition~\rcite{Johansson:2015oia}. Further, we will show how this
decomposition is fixed by factorisation.

\paragraph{$q\bar{q}$ amplitude basis.}

The basis of ordered amplitudes $A(\u{1},\o{2},\sigma)$\footnote{We follow
the notation of \rcite{Johansson:2015oia} for labeling the external particles.
Specifically, we consider all $k$ pairs of quarks and antiquarks
to be of distinct flavours;
for definiteness we use odd underscored labels $\u{1},\u{3},\dots$ for quarks
and even overscored labels $\o{2},\o{4}$ for their respective antiquarks.
The remaining $(n-2k)$ gluons are labeled with plain numbers such as $5$.
Within graphs, we find it convenient to distinguish the flavours of the quark lines
by different colourings.
Amplitudes with multiple quark pairs of the same flavour may be obtained
from equal-mass, distinct-flavour amplitudes 
by summing over permutations of a chosen subset of quark labels,
see \eqn{FlavorPermutation} below.
}
with two like-flavoured quarks fixed is built from
permutations $\sigma$ of the remaining $(n-2)$ labels. These allow gluons in
arbitrary positions but impose more structure on the positions of the quark labels.
A simple way to describe this is that the quark labels must correspond
to legal ``bracket structures''. For definiteness we associate the
quark labels
with opening brackets ``\qo{black}''
and antiquark labels with closing brackets ``\qc{black}''.
Intuitively, a closing bracket must always follow an opened
and previously unclosed bracket, so brackets structures like
$\qo{black} \qc{black} \qc{black} \qo{black} \qo{black} \qc{black}$ are illegal.
All such opening-closed bracket pairs correspond to quark-antiquark pairs. We
can further embellish these bracket labels with colour in order to distinguish
the flavour information.
For instance, we can equate the permutation
$(\u{3},\u{5},\o{6},\u{7},\o{8},\o{4})$
to the bracket structure
$\qo{green} \qo{red} \qc{red} \qo{blue} \qc{blue} \qc{green}$.

Naturally, all bracket structures can be seen as arising from a recursive
construction from shorter bracket structures, that is
\be
   k\text{-bracket} = \qo{black}(i-1)\text{-bracket}\qc{black}
                        \oplus (k-i)\text{-bracket} .
\ee
We can precisely formulate this in a recursive definition for all flavoured
quark brackets. For a given set of flavours $F$, the set of associated bracket
structures is given by
\be
   {\cal Q}_F = \bigcup_{f \in F} \bigcup_{E \in \mathbb{P}(F \setminus f)}
      \big\{ (\substack{ \{ \\ \\ \textstyle f \\ \\ \phantom A }) \oplus \pi
      \oplus (\substack{ \} \\ \\ \textstyle \bar{f} \\ \\ \phantom A}) \oplus
             \rho ~\big|~
             (\pi,\rho) \in {\cal Q}_{E}
                        \times {\cal Q}_{(F\setminus f) \setminus E}
      \big\} , 
\label{QuarkBrackets}
\ee
where $\mathbb{P}(S)$ is the set of all subsets of $S$, or the power set of $S$.
The base of the above recursion is ${\cal Q}_\emptyset$, which contains only an
empty ordering $()$. This formulation is useful as it will allow us to easily
extend to the results of the following sections. Furthermore, it provides a way
to compute how many such brackets there are:
\beal
   \big| {\cal Q}_{2k} \big|
       = \sum_{i=1}^{k} \binom{k}{i} \binom{i}{1}
         \big|{\cal Q}_{2(i-1)}\big| \big|{\cal Q}_{2(k-i)}\big|
       = \frac{(2k)!}{(k+1)!} ,
\label{DyckWordCount}
\eeal
where ${\cal Q}_{2k}$
is a shorthand for the set of distinctly flavoured quark brackets of length~$2k$.

The above notation allows us to formulate Melia's amplitude basis as\footnote{In
contrast to the notation used for the the basis set of permutations associated
to a basis, here we use a similar notation for the basis set of ordered
amplitudes,
${\cal B}^{a,b}_{n,k} = \{ A(a,\chi,b)~|~\chi \in {\cal B}^{a,b}_X\}$,
highlighting the particle and quark-pair counts $n$ and~$k$.
}
\be
   {\cal B}_{n,k}^{2,1} =
   \big\{ A(\o{2},\sigma,\u{1}) ~\big|~
          \sigma \in {\cal Q}_{2(k-1)} \shuffle {\cal G}_{n-2k} \big\},
\label{MeliaBasis}
\ee
where the unfixed quark brackets in ${\cal Q}_{2(k-1)}$
are shuffled\footnote{The standard definition
of the shuffle product $\rho \shuffle \sigma$ of two orderings of length $r$ and $s$
is the set of $(r+s)!/(r!\,s!)$ ways of interleaving them,
\ie permuting $\rho \oplus \sigma$ without changing their internal order.
} with arbitrary permutations of gluon labels, as denoted by
\be
   {\cal G}_{n-2k} = \big\{ \sigma \in {\rm S}_G ~\big|~
      G = \{g_{2k+1},\dots,g_n\} \big\} .
\label{GluonPermutations}
\ee
The size of the basis is then immediately~\cite{Melia:2013epa}
\be
   \big|{\cal B}_{n,k}^{1,2}\big|
    = |{\cal Q}_{2(k-1)}| \times (n-2k)! \times \frac{(n-2)!}{(2k-2)!(n-2k)!}
    = \frac{(n-2)!}{k!} .
\label{MeliaBasisSize}
\ee

Recall that the cyclic symmetry of colour-ordered amplitudes
allows us to freely rewrite $A(\o{2},\sigma,\u{1})$ as $A(\u{1},\o{2},\sigma)$.
For the five-point four-quark amplitude considered in \sec{sec:five}
only the quark pair $\u{3} \leftarrow \o{4}$ is unfixed,
so the single bracket structure $\qo{green} \qc{green}$
is dressed with three gluon insertions and gives the Melia basis of
$A(\u{1},\o{2},5,\u{3},\o{4})$, $A(\u{1},\o{2},\u{3},5,\o{4})$
and $A(\u{1},\o{2},\u{3},\o{4},5)$.

It is worth pointing out that
the quark-arrow convention for each quark pair may be switched at will.
So the Melia basis $\{A(\u{1},\sigma,\o{2})\}$
(used \eg in \rcite{Brown:2016hck})
is equivalent to $\{A(\o{2},\sigma,\u{1})\}$, as well as
their relabelings $\{A(\u{2},\sigma,\o{1})\}$ and $\{A(\o{1},\sigma,\u{2})\}$.
In the specific rendition of the Melia basis above
the arrow of the base quark line $\o{2} \rightarrow \u{1}$ goes
in the opposite way to the rest.
This can be depicted by an outer pair of square brackets.
For instance, the above five-point orderings are
$(\o{2},5,\u{3},\o{4},\u{1}) = \qob{black} 5 \qo{green} \qc{green} \qcb{black}$,
$(\o{2},\u{3},5,\o{4},\u{1}) = \qob{black} \qo{green} 5 \qc{green} \qcb{black}$ and
$(\o{2},\u{3},\o{4},5,\u{1}) = \qob{black} \qo{green} \qc{green} 5 \qcb{black}$.
Square brackets will play a more important role in the subsequent sections,
while at this point they are a matter of convention for the fixed quark pair.

\paragraph{$q\bar{q}$ colour decomposition.}

In order to rewrite the full colour-dressed amplitude ${\cal A}_{n,k}$
in terms of the Melia basis of ordered amplitudes,
we need to characterise
the corresponding colour coefficients $C(\u{1},\o{2},\sigma)$ in the decomposition
\be
   {\cal A}(\u{1}, \o{2}, X)
    =\!\sum_{ \sigma \in {\cal B}^{1,2}_X }\!\!
      C(\u{1},\o{2},\sigma) A(\u{1},\o{2},\sigma) .
\label{JOM}
\ee
They are given by certain vertical-ladder colour diagrams
most easily understood graphically~\cite{Johansson:2015oia}.
For instance, for the $q\bar{q}$ stretch
of the five-point amplitude studied in \sec{sec:five}
this decomposition involves the following colour factors:
\begin{subequations} \begin{align}\!
   C(\u{1},\o{2},5,\u{3},\o{4}) & = C\!\left(\scalegraph{0.86}{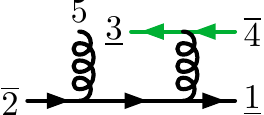}\right) ,
   \qquad \quad
   C(\u{1},\o{2},\u{3},\o{4},5) = C\!\left(\scalegraph{0.86}{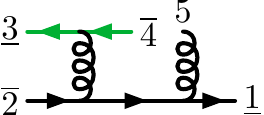}\right) , \\*
   C(\u{1},\o{2},\u{3},5,\o{4}) & = C\!\left(\scalegraph{0.86}{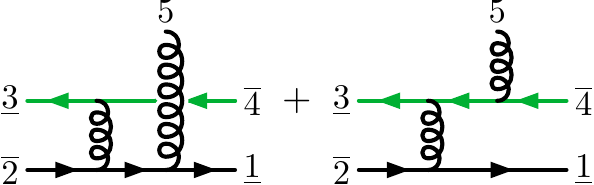}\right) 
    = -T^{a_5}_{i_1 \bar k} T^b_{k \bar\imath_2} T^b_{i_3 \bar\imath_4}
    + T^b_{i_1 \bar\imath_2} T^b_{i_3 \bar k} T^{a_5}_{k \bar\imath_4} .
\label{CqqQgQ}
\end{align} \label{A5qqColor}%
\end{subequations}
Here the last line contains more than one diagram
due to gluon 5 being in a nested position
$\qob{black} \qo{green} 5 \qc{green} \qcb{black}$
--- surrounded by more than one pair of quark brackets.
Moreover, the first two colour coefficients illustrate the fact that
the quark pair $\u{3}\leftarrow\o{4}$ as a whole behaves much like a gluon,
so it easy to imagine that a quark pair in a nested position
with respect to other quarks also generates summation
over different ways to contract its adjoint index (shown by the curly line).

A general construction for the colour coefficients in \eqn{JOM} can be implemented
by the following algebraic formula
\be
   C(\u{1},\o{2},\sigma) = (-1)^{k-1}\,
      \qob{black}2|\sigma|1\qcb{black} {\Bigg| \footnotesize
      \begin{array}{l}
         g~\,\rightarrow~\Xi_{l(g)}^{a_g} \\
         q~\,\rightarrow\,\qo{black}q|\,T^b\!\otimes \Xi_{l(q)-1}^b \\
         \bar{q}~\,\rightarrow~|q\qc{black}
      \end{array} } ,
\label{JOcolor}
\ee
where a bra-ket notation represents the fundamental indices, \eg
\be
   \qo{black}q|T^a|\bar{q}\qc{black} = T^a_{i_q \bar\imath_{\bar{q}}} , \qquad \quad
   \qob{black}2|T^a|1\qcb{black} = T^a_{\bar\imath_2 i_1} = -T^a_{i_1 \bar\imath_2} .
\ee
Then the replacement rules in \eqn{JOcolor}
convert a bracketed permutation $\sigma$
to an expression in terms of the usual generators $T^a$
and new tensor-representation generators
\beal &~\;\quad
   \underset{\substack{~ \\ \rotatebox[origin=c]{60}{=}}}{}\Xi_l^a \,=\, 
      \sum_{r=1}^{l}\,
      \underbrace{\,\mathbb{I} \otimes \cdots \otimes \mathbb{I} \otimes
      \overbrace{T_r^a \otimes \mathbb{I} \otimes \cdots \otimes \mathbb{I}
                     \otimes \o{\mathbb{I}}\,}^{r}
                 }_{l} . \\ &
   \scalegraph{0.77}{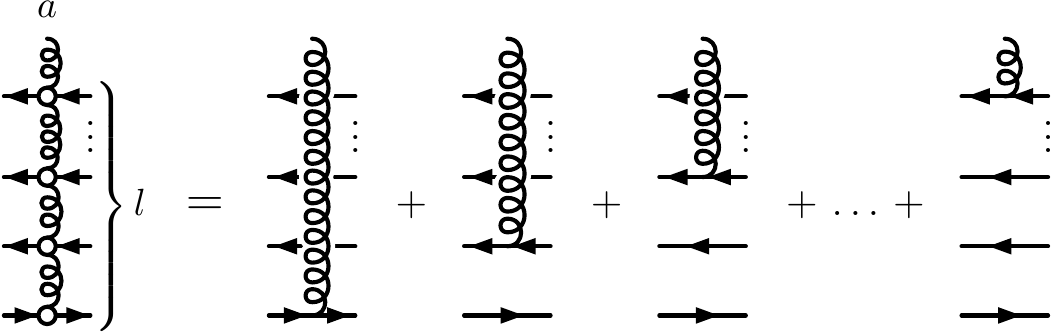}
\label{Xi}
\eeal
They have an explicit dependence on the nestedness level~$l$
of the label being replaced within \eqn{JOcolor}
and encode the summation over different ways to attach an adjoint index
to the quark lines of the surrounding brackets.\footnote{As an explicit example,
we apply the formula~\eqref{JOcolor}
to the non-trivial five-point colour factor in \eqn{A5qqColor}:
\beal
   C(\u{1},\o{2},\u{3},5,\o{4}) &
    = C\qob{black}2, \qo{green}3, 5, 4\qc{green}, 1\qcb{black}
    =-\qob{black}2| \qo{green}3| {\color{green}T}^b\!\otimes\Xi^b_1
      |\Xi^{a_5}_2|4\qc{green} |1\qcb{black}
    =-\qob{black}2| \qo{green}3| {\color{green}T}^b\!\otimes T^b
      |T^{a_5}\!\oplus {\color{green}T}^{a_5}
      |4\qc{green} |1\qcb{black} \\ &
    =-\qob{black}2| T^b T^{a_5} \qo{green}3| {\color{green}T}^b
      |4\qc{green} |1\qcb{black}
    - \qob{black}2| T^b \qo{green}3| {\color{green}T}^b\,{\color{green}T}^{a_5}
      |4\qc{green} |1\qcb{black}
    =-T^b_{\bar\imath_2 k} T^{a_5}_{\bar k i_1} {\color{green}T}^b_{i_3 \bar\imath_4}
    - T^b_{\bar\imath_2 i_1}
      {\color{green}T}^b_{i_3 \bar k} {\color{green}T}^{a_5}_{k \bar\imath_4} .
\eeal
}
The closed-form construction above was
conjectured by Johansson and one of the current authors~\cite{Johansson:2015oia}
and subsequently proven by Melia~\cite{Melia:2015ika}.
Whenever we wish to invoke it,
we will refer to the colour factors~\eqref{JOcolor}
as $C_\text{JO}(\o{2},\sigma,\u{1})$.
Moreover, as any like-flavoured quarks can be exchanged,
the colour factors
\be
   C_\text{JO}(\o{2},\sigma,\u{1}) = C\!\left(\scalegraph{0.86}{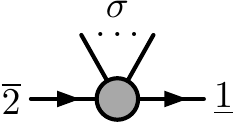}\right) ,
   \qquad \quad
   C_\text{JO}(\u{1},\sigma,\o{2}) = C\!\left(\scalegraph{0.86}{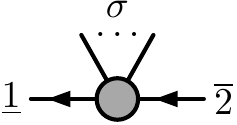}\right)
\ee
differ from each other by a trivial flip of the baseline arrow in \eqn{Xi}.

It is worth noting that although the colour factors
are uniquely specified by the Melia basis~\eqref{MeliaBasis},
they can be expressed in different ways, and the above formulae
give one of many possible colour-diagrammatic representations for them.
For instance, the five-point colour factor~$C(\u{1},\o{2},\u{3},5,\o{4})$
is shown in \eqn{A5qqColor}
as a combination of two diagrams with the adjoint line of gluon~$5$
on the right-hand side of the quark line $\u{3} \leftarrow \o{4}$,
but can also be rewritten with the gluonic line on the left-hand side.

A remarkable and previously unnoticed property of the JO colour
factors is that they indeed satisfy the colour factorisation
relation~\eqref{eq:ColourFactorization} and the leg-exchange
relations~\eqref{CqqFactorization2}--\eqref{CggFactorization}.
These properties indeed follow from the
co-unitarity of the basis to be shown in~\sec{sec:counitarity}.
Alternatively, they can be derived from
the definition of the colour factors~\eqref{JOcolor}
and the algebraic relations~\eqref{JacobiCommutation},
as detailed in \app{sec:qqunitarity}.

\paragraph{$q\bar{q}$ colour from factorisation.}

Let us now present an argument showing that
the co-unitarity of the Melia basis to itself also
completely fixes the JO colour factors in its associated decomposition.
In short, one can systematically apply
either the colour factorisation relation
or the leg-exchange relation until the colour factor is
entirely expressed in terms of three-point colour vertices.
Such a procedure is not unique, reflecting the non-uniqueness
of the closed-form representation for the colour factors.
Nevertheless,
it provides a useful recursive way of implementing these colour factors,
as the only place where any convention may enter
is in the definition of the three-point colour factors.

We will make use of three identities.
The first two are the colour factorisation relation~\eqref{eq:ColourFactorization}
and the leg-exchange relation for a $q\bar{q}$ pair~\eqref{CqqFactorization2}.
As all amplitudes in the Melia basis have a specific orientation of quark brackets,
both of these equations relate $n$-point colour factors to $(n-1)$-point ones.
The third relation we shall employ
is the quark-gluon exchange relation~\eqref{CqgFactorization},
which necessarily involves two terms of the same multiplicity.
For this reason it can be regarded
as the origin of the multiple terms in the JO colour factors.

Our aim is to apply the three relations in such a way
that we can always decrease the multiplicity of the involved diagrams.
This will necessarily result in a multi-parton colour factor
expressed in terms of three-point colour vertices.
We consider each identity in turn.
\begin{enumerate}[leftmargin=\parindent]
\item
We can apply the colour factorisation relation~\eqref{eq:ColourFactorization}
unless the first and last particles of the permutation
are a quark and antiquark of the same flavour,
where flavour conservation makes this impossible.
If it is possible, we apply the factorisation identity
and thus reduce the multiplicity of the involved diagrams.
\item
Otherwise we look to apply the leg-exchange relation~\eqref{CqqFactorization2}
for a $q\bar{q}$ pair.
This is possible if there is an adjacent $q\bar{q}$ pair in the permutation,
in which case we immediately reduce the multiplicity of the involved diagrams.
\item If we cannot find such a pair,
then there must be one or more gluons enclosed by the most nested quark bracket.
We can then repeatedly use the $qg$ exchange relation~\eqref{CqgFactorization}
to move these gluons to the right of all the quarks in the permutation.
This introduces a number of additional terms,
but they involve diagrams of lower of multiplicity.
The remaining terms of the same multiplicity
no longer have all particles enclosed and are therefore again amenable
to the colour factorisation identity.
\end{enumerate}
In this way, we can repeatedly apply the three relations
until we reach an expression for the colour factor
entirely given in terms of three-point vertices.
Notably, we chose to move all gluons to the right,
which corresponds to the non-uniqueness of the JO factors.

To demonstrate the above factorisation procedure,
let us derive the five-point colour factor
presented earlier in \eqn{CqqQgQ}:
\begin{align}
\label{CqqQgQderivation}
 & C\!\left(\scalegraph{0.86}{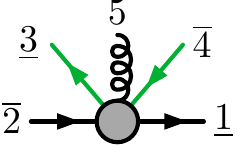}\right)
    = C\!\left(\scalegraph{0.86}{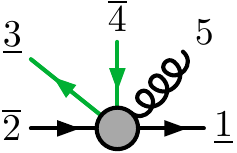}\right)\!
    + C\!\left(\scalegraph{0.86}{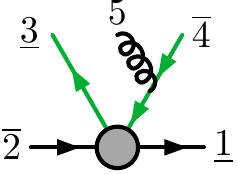}\right) \\ &
    = C\!\left(\scalegraph{0.86}{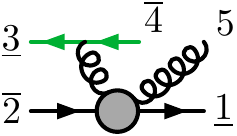}\right)
    + C\!\left(\scalegraph{0.86}{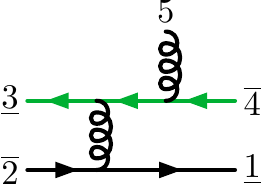}\right)
    = C\!\left(\scalegraph{0.86}{CqqQQg}\right)
    + C\!\left(\scalegraph{0.86}{CqqQgQ2}\right)\!. \nn
\end{align}
Here the starting diagram cannot be directly factorised
via the relations~\eqref{eq:ColourFactorization} or~\eqref{CqqFactorization2},
so in the first line we move gluon~$5$ using
the quark-gluon exchange relation~\eqref{CqgFactorization}.
The resulting additional term contains a three-point vertex
connected to a four-point amplitude sub-diagram,\footnote{In \eqn{CqqQgQderivation}
and below we use a slightly streamlined notation with respect
to \eg eqs.~\eqref{eq:ColourFactorization}--\eqref{CggFactorization}
by allowing both amplitude blob diagrams and explicit vertices
within a single colour-factor diagram.
}
which can then be converted
to an explicit colour factor via the $q\bar{q}$
factorisation identity~\eqref{CqqFactorization2}.
This identity also factorises
the remaining five-point term 
onto a four-point sub-diagram,
which is finally amenable to the basic colour factorisation relation~\eqref{eq:ColourFactorization}.

\subsection{Distinct-flavour stretch}
\label{sec:qQ}

Starting from this section, we introduce new bases of colour-ordered amplitudes
in QCD, as well as the corresponding colour decompositions. We leave a detailed
discussion of their properties to~\sec{sec:counitarity}.

\paragraph{$qQ$ amplitude basis.}

We begin with the case of two distinctly flavoured quarks fixed next
to each other. Without loss of generality, let us label the fixed particles as
quark~$\u{1}$ and antiquark~$\o{4}$.
Within the basis of ordered amplitudes, any ordering then starts with a standard
opening bracket~``\qo{black}'' and ends with a standard closing
bracket~``\qc{green}''. The consistency of the bracket structure implies that
the former must be closed with ``\qc{black}''$\Leftrightarrow\o{2}$
before the latter is opened with ``\qo{green}''$\Leftrightarrow\u{3}$.
So in contrast to the $q\bar{q}$ case, the particles appearing in-between
are not automatically enclosed by any external brackets.
Now in order to build a set of permutations that corresponds to an amplitude basis,
one needs to introduce a new feature to our bracket structures ---
all unenclosed quark brackets should be allowed
to appear in both kinds of arrow orientations.
More enclosed quark pairs, however, should only come with one of two orderings,
which we here choose to be the canonical one. We note that this is indeed a
choice of a linearly independent set of amplitudes, corresponding to a given
``signature'' of quark line orientation~\cite{Melia:2013epa}. 
For instance, brackets structures like
$\qo{black}\qc{black} \qo{red} \qo{blue}\qc{blue} \qc{red} \qo{green}\qc{green}$,
$\qo{black}\qc{black} \qob{red} \qo{blue}\qc{blue} \qcb{red} \qo{green}\qc{green}$,
$\qo{black}\qc{black} \qob{red} \qcb{red} \qob{blue}\qcb{blue} \qo{green}\qc{green}$
or
$\qo{black} \qo{blue}\qc{blue} \qc{black} \qob{red} \qcb{red} \qo{green}\qc{green}$
are now all legal, but
$\qo{black}\qc{black} \qo{red} \qob{blue}\qcb{blue} \qc{red} \qo{green}\qc{green}$,
$\qo{black}\qc{black} \qob{red} \qob{blue}\qcb{blue} \qcb{red} \qo{green}\qc{green}$
or
$\qo{black} \qob{blue}\qcb{blue} \qc{black} \qob{red} \qcb{red} \qo{green}\qc{green}$
are not.
This pattern is consistent with the observations of \rcite{Kalin:2017oqr}
at one loop.
Note that this feature does not arise until six-point amplitudes, as a $qQ$ basis
for the five-point amplitude considered in \sec{sec:five} can be shown to be
$\qo{black} 5 \qc{black} \qo{green}\qc{green}$,
$\qo{black}\qc{black} 5 \qo{green}\qc{green}$ and
$\qo{black}\qc{black} \qo{green} 5 \qc{green}$.
To be more illustrative of the new features of the basis,
let us present a six-quark basis in the context of a complete colour decomposition
\beal\!\!\!\!
   {\cal A}^{\text{tree}}_{6,3}\!
    = C\!\left(\scalegraph{0.8}{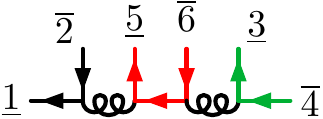}\right)\!
      \substack{ ~~\,\qo{black}~\;\qc{black}~\;
                 \qo{red}~\;\qc{red}~\;\qo{green}~\;\qc{green} \\
                 \textstyle A(\u{1},\o{2},\u{5},\o{6},\u{3},\o{4}) \\ \\ \phantom A
               }
    + C\!\left(\scalegraph{0.8}{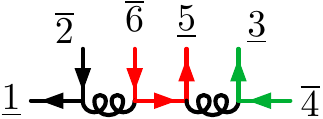}\right) &
      \substack{ ~~\,\qo{black}~\;\qc{black}~\;
                 \qob{red}~\;\qcb{red}~\;\qo{green}~\;\qc{green} \\
                 \textstyle A(\u{1},\o{2},\o{6},\u{5},\u{3},\o{4}) \\ \\ \phantom A
               } \!\!\!\! \\
    + C\!\left(\scalegraph{0.8}{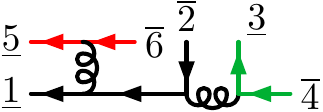}\right)\!
      \substack{ ~~\,\qo{black}~\;\qo{red}~\;\qc{red}~\;\qc{black}~\;
                 \qo{green}~\;\qc{green} \\
                 \textstyle A(\u{1},\u{5},\o{6},\o{2},\u{3},\o{4}) \\ \\ \phantom A
               }
    + C\!\left(\scalegraph{0.8}{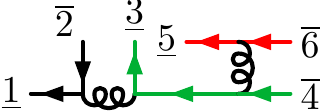}\right) &
      \substack{ ~~\,\qo{black}~\;\qc{black}~\;
                 \qo{green}~\;\qo{red}~\;\qc{red}~\;\qc{green} \\
                 \textstyle A(\u{1},\o{2},\u{3},\u{5},\o{6},\o{4}) \\ \\ \phantom A
               } .\!\!\!\!\!
\eeal
The form of the colour factors here can be derived from the principal unitarity
constraint~\eqref{eq:ColourFactorization} and the five-point colour
factors~\eqref{A5qgColorFactorization}. Unlike in Melia's basis, the
$qQ$-stretched base line is now separated into multiple segments, the leftmost
and rightmost being $\u{1} \leftarrow \o{2}$ and $\u{3} \leftarrow \o{4}$ with
possible other particles in-between.

Intuitively, all $k$-bracket structures with both orientations of unenclosed
brackets can be divided into its first such bracket with only unflipped brackets
inside and a similar remainder,
\begin{equation}
\label{DyckWordCountWithFlips1}
   k\text{-bracket w\:\!\!/ flips} = \left\{
   \begin{aligned}
   \text{either }\qo{black}(i-1)\text{-bracket}\qc{black}&
      ~\oplus~(k-i)\text{-bracket w\:\!\!/ flips} \\
   \text{or }\:\qob{black}(i-1)\text{-bracket}\qcb{black}&
      ~\oplus~(k-i)\text{-bracket w\:\!\!/ flips} ,
   \end{aligned} \right. \\
\end{equation}
We can precisely capture this behaviour by introducing a set of all quark brackets
with flips allowed for those in an unenclosed position:
\beal
   \o{\cal Q}_F = \bigcup_{f \in F} \bigcup_{E \in \mathbb{P}(F \setminus f)} &
      \big\{ (\substack{ \{ \\ \\ \textstyle f \\ \\ \phantom A }) \oplus \pi
      \oplus (\substack{ \} \\ \\ \textstyle \bar{f} \\ \\ \phantom A}) \oplus
             \rho ~\big|~
             (\pi,\rho) \in {\cal Q}_{E}
                        \times \o{\cal Q}_{(F\setminus f) \setminus E}
      \big\} \\ \cup~&
      \big\{ (\substack{ [ \\ \\ \textstyle \bar{f} \\ \\ \phantom A }) \oplus \pi
      \oplus (\substack{ ] \\ \\ \textstyle f \\ \\ \phantom A}) \oplus
             \rho ~\big|~
             (\pi,\rho) \in {\cal Q}_{E}
                        \times \o{\cal Q}_{(F\setminus f) \setminus E}
      \big\}, 
\label{QuarkBracketsWithFlips}
\eeal
where each time an enclosed bracket occurs,
it is taken from ${\cal Q}_F$ defined previously in \eqn{QuarkBrackets}.
Again, the base of the recursion is that
$\o{\cal Q}_\emptyset$ is a set with an empty ordering.
The recursive nature of this definition 
allows us to easily obtain a counting for such structures:
\begin{align}
\label{DyckWordCountWithFlips2}
   \big|\o{\cal Q}_{2k}\big| &
    = \sum_{i=1}^{k} 2 \binom{k}{i} \binom{i}{1}
      \big|{\cal Q}_{2(i-1)}\big| \big|\o{\cal Q}_{2(k-i)}\big|
    = \frac{(2k)!}{k!} ,
\end{align}
where the factor of two accounts for the canonical and flipped bracket
configurations of the first bracket pair comprising $(i-1)$ standard brackets
and followed by the remaining bracket structures with flips.

In terms of the above set, we define the new basis as
\beal
   {\cal B}_{n,k}^{1,4} =
   \big\{ A(\u{1},\sigma,\o{4}) ~\big|~
          (\u{1})\oplus\sigma\oplus(\o{4})\,
          \in \o{\cal Q}_{2k} \shuffle {\cal G}_{n-2k} \big\}
\label{qQBasis}
\eeal
where $k>2$ and the concatenation~$(\u{1})\oplus\sigma\oplus(\o{4})$
explicitly selects only those elements of~$\o{\cal Q}_{2k}$
that begin with $\u{1}$ and end with $\o{4}$.
We note that whilst the set is too large to describe the present basis, this
is because it unifies the treatment of all bases described in this paper.
Let us now check that for the basis~\eqref{qQBasis} its size equals that of the
Melia basis~\eqref{MeliaBasis}. For that, we construct a pure-quark bracket as
\be
   \qo{black}(i-1)\text{-bracket}\qc{black}
      ~\oplus~(j-i)\text{-bracket w\:\!\!/ flips}
      ~\oplus~\qo{green}(k-j-1)\text{-bracket}\qc{green} .
\ee
Its counting is then computed as
\be
   \big|{\cal B}_{n=2k,k}^{1,4}\big| = \sum_{i=1}^{k-1} \sum_{j=i}^{k-1}
      \binom{k-2}{i-1} \binom{k-i-1}{k-j-1}
      \big|{\cal Q}_{2(i-1)}\big| \big|\o{\cal Q}_{2(j-i)}\big|
      \big|{\cal Q}_{2(k-j-1)}\big|
    = \frac{(2k-2)!}{k!} .
\ee
This is indeed $\big|{\cal Q}_{2(k-1)}\big|$,
so dressing the quark bracket structures with gluons
promotes it to the required $(n-2)!/k!$ analogously to \eqn{MeliaBasisSize}.

\paragraph{$qQ$ colour decomposition.}

Let us discuss the colour factors in the decomposition onto the distinct-flavour basis
\be
   {\cal A}(\u{1}, X, \o{4}) 
    =\!\sum_{\sigma \in {\cal B}_X^{1,4}}\!
      C(\u{1},\sigma,\o{4}) A(\u{1},\sigma,\o{4}) .
\label{qQdecomposition}
\ee
In order to formulate a closed-form expression for the colour factors,
we note that any given permutation $\sigma$
is naturally split by the quarks in the unnested positions
into $(2u-1)$ sub-permutations, \eg
\be
   (\u{1},\sigma,\o{4}) = \big(
      \qo{black}1,\sigma_1,2\qc{black},\sigma_2,
      \qo{red}5,\sigma_3,6\qc{red},\sigma_4,\dots,
      \sigma_{2u-2},
      \qo{green}3,\sigma_{2u-1},4\qc{green} \big) , \qquad \quad 2 \leq u \leq k ,
\label{qQpermutation}
\ee
where the evenly numbered $\sigma_{2v}$ are defined to be purely gluonic.
Since all permutations in the $qQ$ basis~\eqref{qQBasis}
can be obtained from this configuration by relabelling,
writing its colour factor is sufficient
to specify the complete decomposition~\eqref{qQdecomposition}.
We can now use this breakdown of the permutation to find a closed form for the
colour factor in terms of previously defined building blocks. By construction,
the colour factorisation formula~\eqref{eq:ColourFactorization} allows us to write
\begin{subequations}
\be
   C\!\left(\scalegraph{0.86}{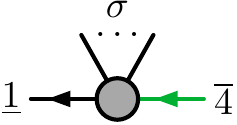}\!\right)
    = C\!\left(\scalegraph{0.86}{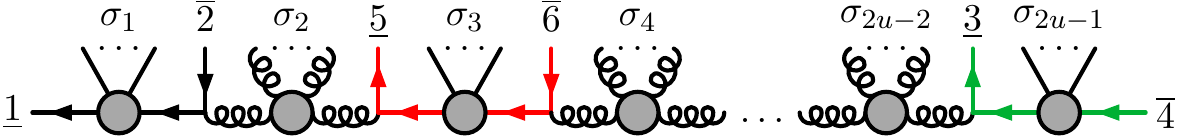}\right) ,
\label{qQdiagrams}
\ee
which translates to
\begin{align}
 & C(\u{1},\sigma,\o{4})_{\imath_1 \bar\imath_4} \nn
    = C_\text{JO}(\u{1},\sigma_1,\o{2})_{\imath_1 \bar\jmath_2}
      T^{b_1}_{\jmath_2 \bar\imath_2}
      C_\text{DDM}(g_1,\sigma_2,g_2)^{b_1 b_2} \bigg. \nn \\* & \qquad \times\!
      \prod_{v=3}^{u} \Big\{ T^{b_{2v-4}}_{\imath_{2v-1} \bar\jmath_{2v-1}}
      C_\text{JO}(\u{2v\!-\!1},\sigma_{2v-3},\o{2v}
                  )_{\jmath_{2v-1} \bar\jmath_{2v}}
      T^{b_{2v-3}}_{\jmath_{2v} \bar\imath_{2v}} \\* &
      \qquad \qquad \qquad \qquad\;\,\times
      C_\text{DDM}(g_{2v-3},\sigma_{2v-2},g_{2v-2})^{b_{2v-3} b_{2v-2}}
      \Big\}\,
      T^{b_{2u-2}}_{\imath_3 \bar\jmath_3}
      C_\text{JO}(\u{3},\sigma_{2u-1},\o{4})_{\jmath_3 \bar\imath_4} . \nn
\end{align} \label{qQcolor}%
\end{subequations}
Here we have only made explicit the colour indices relevant for the factorisation.
Recall that the $q\bar{q}$-stretch colour factors have been
introduced in \eqn{JOcolor},
whereas the DDM colour factors are given by strings of structure constants, namely
\be
   C_\text{DDM}(1,\sigma,n) = C\!\left(\scalegraph{0.86}{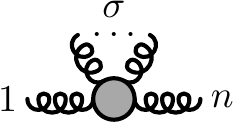}\right)
    = \tf^{\,a_1 a_{\sigma(2)} b_1} \tf^{\,b_1 a_{\sigma(3)} b_2} \dots
      \tf^{\,b_{n-3} a_{\sigma(n-1)} a_n} .
\label{DDMcolor}
\ee
We supplement it with the trivial extension
$C_\text{DDM}(1,2)^{a_1 a_2} = \delta^{a_1 a_2}$
to account for the case where a purely gluonic permutation $\sigma_{2v}$
in \eqref{qQcolor} happens to be empty.

As an amusing switch of perspective,
one could now regard the $q\bar{q}$ colour factors~\eqref{JOcolor}
as a particular case of the above $qQ$ colour factors,
for which no initial factorisation into flavour-neutral blocks,
as in \eqn{qQdiagrams}, is possible.

\subsection{Stretching with gluons}
\label{sec:qg}

In this section we present the amplitude bases
and the corresponding colour decompositions
for the cases where one or both fixed particles are gluons.

\paragraph{Quark-gluon stretch.}
Without loss of generality, we take the fixed quark and gluon to be $\u{1}$ and $n$.
Then the new basis is defined as
\beal
   {\cal B}_{n,k}^{1,n} =
   \big\{ A(\u{1},\sigma,n) ~\big|~
          (\u{1})\oplus\sigma\,\in \o{\cal Q}_{2k} \shuffle {\cal G}_{n-2k} \big\}
\label{qgBasis}
\eeal
where $n>2k$.
Similarly to the $qQ$ basis~\eqref{qQBasis},
the quark lines that appear in an unnested position to right of label $\o{2}$
may come in two orientations.
Conveniently, we have already considered a five-point example of such a basis
in \sec{sec:five}, and in the bracket notation it is given by
$\qo{black}\qc{black} \qo{green}\qc{green} 5$,
$\qo{black}\qc{black} \qob{green}\qcb{green} 5$ and
$\qo{black} \qo{green}\qc{green} \qc{black} 5$.
As the basis is constructed from the same set of quark brackets as the $qQ$
basis, the same rules about nesting of square brackets apply. For example, if we
consider a seven-point six-quark basis this involves orderings corresponding to
$\qo{black}\qc{black} \qob{green} \qo{red} \qc{red} \qcb{green} 7$ and
$\qo{black}\qc{black} \qob{red}\qcb{red} \qob{green}\qcb{green} 7$,  but not
$\qo{black}\qc{black} \qob{green} \qob{red} \qcb{red} \qcb{green} 7$.

There are multiple ways to define the amplitude decomposition into the $qg$ basis,
\be
\label{qgDecomposition}
   {\cal A}(\u{1}, X, n)
    =\!\sum_{\sigma \in {\cal B}_X^{1,n}}\!
      C(\u{1},\sigma,n) A(\u{1},\sigma,n) .
\ee
Here we choose to define it by factorising onto the $qQ$ colour factors~\eqref{qQcolor}.
In other words, since every permutation~$\sigma$ may be split into two
by the rightmost-occurring quark, which we denote as $\bar{q}$,
we depict the $qg$ colour factors as
\begin{subequations}
\be
   C\!\left(\scalegraph{0.86}{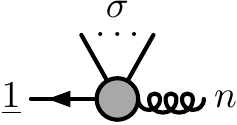}\right)
    = C\!\left(\scalegraph{0.86}{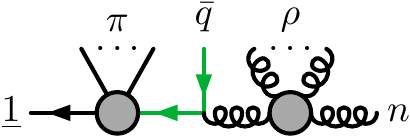}\right) , \qquad \quad
   \sigma = \big(\pi,\bar{q}\qc{green},\rho\big) ,
\ee
More explicitly, we write
\be
   C(\u{1},\sigma,n)_{\imath_1}^{~\,a_n}
    = C(\u{1},\pi,\bar{q})_{\imath_1\bar\jmath}
      \big[{-}T^b_{\jmath\:\!\bar\imath_{\bar{q}}} \big]
      C_\text{DDM}(g,\rho,n)^{b\:\!a_n} .
\ee \label{qgColor}%
\end{subequations}
In the particular case where the rightmost quark $\bar{q}$ is $\o{2}$,
the distinct-flavour colour factor $C(\u{1},\pi,\bar{q})$ should be replaced
by the like-flavour one~\eqref{JOcolor},
which can otherwise be considered as its degeneration.

\paragraph{Gluon-gluon stretch.}
Now fixing two gluons next to each other
and labelling them as $(n-1)$ and $(n-2)$,
we define a new basis and the corresponding colour decomposition as
\beal
   {\cal B}_{n,k}^{n-1,n} & =
   \big\{ A(n\!-\!1,\sigma,n) ~\big|~
          \sigma \in \o{\cal Q}_{2k} \shuffle {\cal G}_{n-2k-2} \big\} , \\
   {\cal A}(n\!-\!1, X, n) &
    =\!\sum_{\sigma \in {\cal B}_X^{n-1,n}}\!
      C(n\!-\!1,\sigma,n) A(n\!-\!1,\sigma,n) ,
\label{ggBasis}
\eeal
where $n \geq 2k+2$.
Since no quark label is fixed,
we should now allow all unenclosed quark brackets to appear in both orientations,
while brackets in nested positions are chosen in the standard orientation.
For example,
a six-parton amplitude with two gluons can be expanded in the basis of
\beal
 & 5 \qo{black}\qc{black} \qo{green}\qc{green} 6 , \qquad \quad
   5 \qo{black}\qc{black} \qob{green}\qcb{green} 6 , \qquad \quad
   5 \qob{black}\qcb{black} \qo{green}\qc{green} 6 , \qquad \quad
   5 \qob{black}\qcb{black} \qob{green}\qcb{green} 6 , \\
 & 5 \qo{green}\qc{green} \qo{black}\qc{black} 6 , \qquad \quad
   5 \qo{green}\qc{green} \qob{black}\qcb{black} 6 , \qquad \quad
   5 \qob{green}\qcb{green} \qo{black}\qc{black} 6 , \qquad \quad
   5 \qob{green}\qcb{green} \qob{black}\qcb{black} 6 , \\
 & 5 \qo{black} \qo{green}\qc{green} \qc{black} 6 , \qquad \quad
   5 \qob{black} \qo{green}\qc{green} \qcb{black} 6 , \qquad~\:\:
   5 \qo{green} \qo{black}\qc{black} \qc{green} 6 , \qquad~\:\:
   5 \qob{green} \qo{black}\qc{black} \qcb{green} 6 .
\eeal

There are many ways to formulate the $gg$-stretch colour decomposition.
For instance, finding the leftmost-occuring quark~$q$,
we can split a DDM building block
off the $qg$-stretch colour factor defined in \eqn{qgColor} above:
\begin{subequations}
\be
   C\!\left(\scalegraph{0.86}{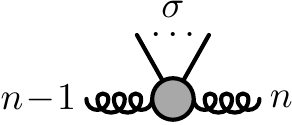}\right)
    = C\!\left(\scalegraph{0.86}{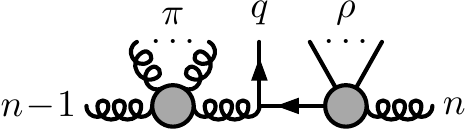}\right) , \qquad \quad
   \sigma = \big(\pi,\qo{black}q,\rho\big) .
\ee
In other words, we have
\be
   C(n\!-\!1,\sigma,n)^{a_{n-1} a_n}
    = C_\text{DDM}(n\!-\!1,\pi,g)^{a_{n-1} b}
      \big[{-}T^b_{\imath_q \bar\jmath} \big]
      C(\u{1},\rho,n)_\jmath^{~\,a_n} .
\ee \label{ggColor}%
\end{subequations}

\subsection{Properties of tree-level bases}
\label{sec:properties}

In this section we demonstrate the crucial properties of the new amplitude bases,
on which the validity and mutual consistency of their colour decompositions rely.

\subsubsection{Co-unitarity}
\label{sec:counitarity}

Here we use the definitions of the bases above to show
that they satisfy the co-unitarity property~\eqref{CoUnitarity}
in the factorisation channel of the form $s_{1P} = s_{nR} \to 0$,
where $P$ and $R$ are non-empty particles sets covering the unfixed labels,
$P \cup R = \{2,\dots,n-1\}$.
Co-unitarity of the bases is crucial
for the subsequent proof of their validity as bases,
as well as the ensuing colour factorisation formulae,
from which any colour decomposition can be derived.

There are two special amplitude bases that
in all limits of the form $s_{1P} \to 0$ factorise onto themselves,
\ie their lower-point versions.
These are the KK basis~\cite{Kleiss:1988ne} in the purely gluonic case,
and the Melia basis~\cite{Melia:2013bta,Melia:2013epa} in presence of quarks.

\paragraph{Purely gluonic case.}

In our notation, the KK basis is
\be
   {\cal B}_{n,0}^{1,n} =
   \big\{ A(1,\sigma,n) ~\big|~ \sigma \in {\cal B}_{P \cup R}^{1,n} \big\} ,
   \qquad \quad
   {\cal B}_{P \cup R}^{1,n} = {\rm S}_{P \cup R} .
\label{KKbasis}
\ee
In the purely gluonic case
there is a non-vanishing residue $\Res_{s_{1P}=0} A(1,\pi,\rho,n)$
for any two suborderings $\pi\in{\rm S}_P$ and $\rho\in{\rm S}_R$.
Therefore, the definition~\eqref{BasisFactorization} implies that
the set of factorised suborderings is
\be
   {\cal U}_{P,R}\big[{\cal B}_{P \cup R}^{1,n}\big]
    = \big\{ (\pi,\rho) \in {\rm S}_P \times {\rm S}_R \big\}
    = {\cal B}_{P}^{1,\bar{p}} \times {\cal B}_{R}^{p,n} ,
\label{CoUnitarityKK}
\ee
so the co-unitarity property is manifest.

\paragraph{Melia-basis case.}

Now let us consider when the factorisation limit $s_{qP} = s_{\bar{q}R} \to 0$
of an ordered amplitude in the Melia basis~\eqref{MeliaBasis}
gives a non-zero residue.
Since quarks~$q$ and~$\bar{q}$ are fixed to be on different sides of the limit,
flavour conservation demands that for there to be a non-zero limit, the
factorisation channel must be of this flavour and therefore that no
other quark
may enter $P$ unless its antiquark is also in $P$.
In terms of bracket structures,
this means that factorisation channels of the form
\be
   \Res_{s_{qP}=0} A(\qob{black}\!
      \underbrace{\dots\!\qo{green}\!\dots}_{\pi \in {\rm S}_P}
      \big|
      \underbrace{\dots\!\qc{green}\!\dots}_{\rho \in {\rm S}_R}\!
      \qcb{black}) = 0
\ee
must vanish, as opposed to
\be
   \Res_{s_{qP}=0} A(\qob{black}\!
   \underbrace{\dots\!\qo{green}\!\dots\!\qc{green}\!\dots}_{\pi \in {\rm S}_P}
   \big|
   \underbrace{\dots\!\qo{red}\!\dots\!\qc{red}\!\dots}_{\rho \in {\rm S}_R}\!
   \qcb{black}) = i
   A(\qob{black}\!\dots\!\qo{green}\!\dots\!\qc{green}\!\dots\!\qcb{black})
   A(\qob{black}\!\dots\!\qo{red}\!\dots\!\qc{red}\!\dots\!\qcb{black}) \neq 0 ,
\ee
where we have indicated the left-right separation between $P$ and $R$ by a
vertical line.
Therefore, denoting the quark flavours entirely in $P$ by $F_P$
and those in $R$ by $F_R$, we see no overlap between these two sets.
The requisite bracket structures are ${\cal Q}(F_P)$ and ${\cal Q}(F_R)$.
As ${\cal Q}(F)$ contains all bracket structures involving all flavours,
those that survive the limit must necessarily be built from ${\cal Q}(F_P)$
and ${\cal Q}(F_R)$.
Using a similar notation for the gluonic sets, $G_P \cup G_R = G =
\{g_{2k+1},\dots,g_n\}$, we can rewrite the relevant Melia bases as
\be
   {\cal B}_{P \cup R}^{q,\bar{q}}
    = {\cal Q}(F_P\!\cup\!F_R) \shuffle S_{G_P \cup G_R} , \qquad \quad
   {\cal B}_{P}^{q,\bar{q}^*}\!\!= {\cal Q}(F_P) \shuffle S_{G_P} , \qquad \quad
   {\cal B}_{R}^{q^*\!,\bar{q}} = {\cal Q}(F_R) \shuffle S_{G_R} ,
\ee
for which from the definition~\eqref{BasisFactorization} we obtain
\begin{align}
   {\cal U}_{P,R}\big[{\cal B}_{P \cup R}^{q,\bar{q}}\big] &
    = \Big\{ (\pi,\rho) \in {\rm S}_P \times {\rm S}_R ~\Big|~
             \pi\oplus\rho \in
             {\cal Q}(F_P\!\cup\!F_R) \shuffle S_{G_P \cup G_R} ,~\!
             \Res_{s_{qP}=0} A(q,\pi,\rho,\bar{q}) \neq 0 \Big\} \nn \\ &
    = \Big\{ (\pi,\rho) ~\Big|~ \pi \in {\cal Q}(F_P) \shuffle S_{G_P} , \quad
             \rho \in {\cal Q}(F_R) \shuffle S_{G_R} \Big\}
    = {\cal B}_{P}^{q,\bar{q}^*}\!\!\times {\cal B}_{R}^{q^*\!,\bar{q}} .
\label{qqBasisFactorization}
\end{align}

\paragraph{New cases.}

There is a significant difference between
three amplitude bases proposed in this paper
and the Kleiss-Kuijf and Melia bases discussed above.
Namely, factorisation limits of the form $s_{1P} \to 0$
intertwine the new bases with one another, as well as with the KK and Melia ones.
In this way, the latter two are the basic cases ---
for this reason we considered them first.
The co-unitarity property of one of the new bases
depends on all of them being co-unitary at the same time.

First, we rewrite the bases as
\begin{subequations} \begin{align}
\label{qQbasis}
   {\cal B}^{q,\bar{Q}}_{P \cup R} &
    = \big\{ \sigma \in {\rm S}_{P \cup R} ~\big|~
             (q)\oplus\sigma\oplus(\bar{Q})\,\in
             \o{\cal Q}(F_P\!\cup\!F_R) \shuffle S_{G_P \cup G_R}
      \big\} , \\
\label{qgbasis}
   {\cal B}^{q,g}_{P \cup R} &
    = \big\{ \sigma \in {\rm S}_{P \cup R} ~\big|~
             (q)\oplus\sigma\,\in
             \o{\cal Q}(F_P\!\cup\!F_R) \shuffle S_{G_P \cup G_R}
      \big\} , \\
\label{ggbasis}
   {\cal B}^{g,g'}_{P \cup R} &
    = \o{\cal Q}(F_P\!\cup\!F_R) \shuffle S_{G_P \cup G_R} ,
\end{align} \label{NewBases}%
\end{subequations}
where $\o{\cal Q} (F)$ allows both bracket orientations
given that they are not enclosed. 
For concreteness,
let us start with that corresponding to a $gg$ stretch in presence of quarks.
We wish to evaluate
\be
   {\cal U}_{P,R}\big[{\cal B}_{P \cup R}^{g,g'}\big]
    = \Big\{ (\pi,\rho) \in {\rm S}_P \times {\rm S}_R ~\Big|~
             \pi\oplus\rho \in {\cal B}_{P \cup R}^{g,g'} ,~\!
             \Res_{s_{gP}=0} A(g,\pi,\rho,g') \neq 0 \Big\} .
\ee

\begin{figure}[t]
\begin{align*}
\text{(a)}~~\scalegraph{0.86}{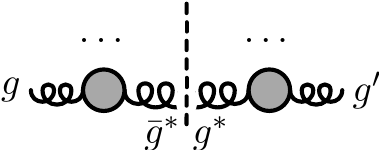} \qquad \quad
\text{(b)}~~\scalegraph{0.86}{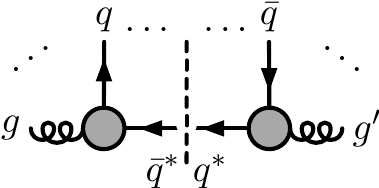} \qquad \quad
\text{(c)}~~\scalegraph{0.86}{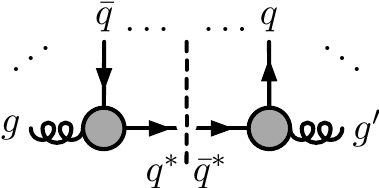}
\end{align*}
\vspace{-15pt}
\caption{\small Factorisation channels of $gg$-basis colour-ordered amplitude}
\label{fig:ggFactorization}
\end{figure}

As shown in \fig{fig:ggFactorization},
there are three ways to have non-zero residues in the $s_{gP}$ channel: the
factorisation channel may have the quantum numbers of a gluon, a quark or an
antiquark, the latter being permitted by a reverse orientation bracket
structure.
We consider them in turn
\begin{enumerate}[label={\alph*)},leftmargin=\parindent]
\item
A non-zero gluonic factorisation channel $s_{gP}$ requires all quarks inside $P$
to be accompanied with the antiquarks of the same flavour, and likewise inside $R$,
which guarantees no overlap between
the bracket structures ${\cal Q}(F_P)$ and ${\cal Q}(F_R)$
without any further constraints.
Hence, in this channel the $gg$ basis
naturally factorises onto two lower-point versions of itself,
\be
   {\cal U}_{P,R}\big[{\cal B}_{P \cup R}^{g,g'}\big]
    = \Big\{ (\pi,\rho) ~\Big|~~
             \pi \in \o{\cal Q}(F_P) \shuffle S_{G_P} ,~~
             \rho \in \o{\cal Q}(F_R) \shuffle S_{G_R} \Big\}
    = {\cal B}_{P}^{g,\bar{g}^*}\!\!\times {\cal B}_{R}^{g^*\!,\bar{g}} ,
\ee
one of which in absence of quarks in $P$ or $R$ may also be the KK basis.

\item
A valid quark factorisation channel $s_{qP}=s_{\bar{q}R}$ needs
all but one quark pair to be entirely inside $P$ or $R$.
Moreover, any quark lines that appear after $q$ in $\pi\in S_P$
or before $\bar{q}$ in $\rho\in S_R$ must enter in a canonical ordering,
as they are enclosed in the $q\bar{q}$ bracket.
This results in an exact factorisation of the $gg$ basis onto two $qg$ bases:
\begin{align}
   {\cal U}_{P,R}\big[{\cal B}_{P \cup R}^{g,g'}\big] &
    = \Big\{ (\pi,\rho) ~\Big| \quad
             \pi\oplus(\bar{q}^*)\,\in \o{\cal Q}(F_P) \shuffle S_{G_P} , \quad
             (q^*)\oplus\rho\,\in \o{\cal Q}(F_R) \shuffle S_{G_R} \Big\} \nn \\ &
    = {\cal B}_{P}^{g,\bar{q}^*}\!\!\times {\cal B}_{R}^{q^*\!,g'} ,
\label{ggBasisFactorization}
\end{align}
where ${\cal B}_{P}^{g,\bar{q}^*}\!$ is a flipped version of \eqn{qgbasis}.

\item
We remind the reader that we use the words ``quark'' and ``antiquark''
merely to label two dual representations of the gauge group
(within the chosen representation labeled by ``flavour''),
and they can be exchanged at will within each pair.
Therefore, the antiquark factorisation channel $s_{\bar{q}P}=s_{qR}$ gives
$~{\cal U}_{P,R}\big[{\cal B}_{P \cup R}^{g,g'}\big]
= {\cal B}_{P}^{g,q^*}\!\!\times {\cal B}_{R}^{\bar{q}^*\!,g'}~$
simply by relabeling the quark channel above.
\end{enumerate}

\begin{figure}[t]
\begin{align*}
\text{(a)} &~~\scalegraph{0.86}{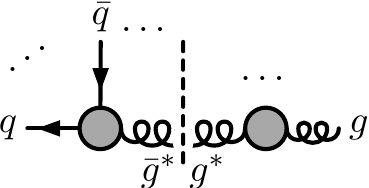} \qquad~\:\quad
\text{(b)}~~\scalegraph{0.86}{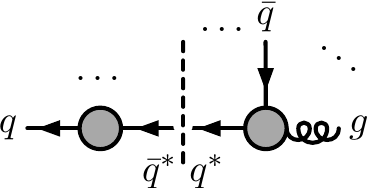} \\
\text{(c)} &~~\scalegraph{0.86}{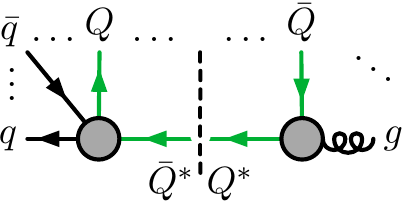} \qquad \quad
\text{(d)}~~\scalegraph{0.86}{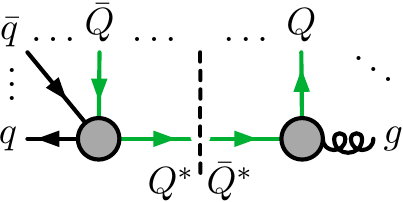}
\end{align*}
\vspace{-15pt}
\caption{\small Factorisation channels of $qg$-basis colour-ordered amplitude}
\label{fig:qgFactorization}
\end{figure}

Next we consider the factorised suborderings~\eqref{BasisFactorization}
of a $qg$ basis.
\Fig{fig:qgFactorization} shows its four non-vanishing channels
of the form $s_{qP}=s_{gR}$:
a gluon, a quark of the same flavour as $q$,
and a quark or an antiquark of another flavour.
Again, the bracket structures govern the orderings
that give a non-zero residue such that
all lower-point combinations are consistently obtained in the limit:
\small
\begin{subequations} \begin{align}
\label{qgBasisFactorization1}
   {\cal U}_{P,R}\big[{\cal B}_{P \cup R}^{q,g}\big] &
    = \Big\{ (\pi,\rho) ~\Big|~
             (q)\oplus\pi\,\in \o{\cal Q}(F_P) \shuffle S_{G_P} , \quad
             \rho \in \o{\cal Q}(F_R) \shuffle S_{G_R} \Big\}
    = {\cal B}_{P}^{q,\bar{g}^*}\!\!\times {\cal B}_{R}^{g^*\!,g} , \\
\label{qgBasisFactorization2}
   {\cal U}_{P,R}\big[{\cal B}_{P \cup R}^{q,g}\big] &
    = \Big\{ (\pi,\rho) ~\Big|~
             \pi \in {\cal Q}(F_P) \shuffle S_{G_P} , \quad
             (q^*)\oplus\rho\,\in \o{\cal Q}(F_R) \shuffle S_{G_R} \Big\}
    = {\cal B}_{P}^{q,\bar{q}^*}\!\!\times {\cal B}_{R}^{q^*\!,g} , \\
\label{qgBasisFactorization3}
   {\cal U}_{P,R}\big[{\cal B}_{P \cup R}^{q,g}\big] &
    = \Big\{ (\pi,\rho) ~\Big|~
      (q)\oplus\pi\oplus(\bar{Q}^*)~\in \o{\cal Q}(F_P) \shuffle S_{G_P} , \quad
      (Q^*)\oplus\rho\,\in \o{\cal Q}(F_R) \shuffle S_{G_R} \Big\}
      \nn \\* &
    = {\cal B}_{P}^{q,\bar{Q}^*}\!\!\times {\cal B}_{R}^{Q^*\!,g} , \\
\label{qgBasisFactorization4}
   {\cal U}_{P,R}\big[{\cal B}_{P \cup R}^{q,g}\big] &
    = {\cal B}_{P}^{q,Q^*}\!\!\times {\cal B}_{R}^{\bar{Q}^*\!,g} .
\end{align} \label{qgBasisFactorization}%
\end{subequations}
\normalsize
In particular, note how in the $q$ channel~\eqref{qgBasisFactorization2}
flavour conservation constrained the left permutation $\pi$
to bracket structures with no flips, thus giving Melia's $qq$ basis
on one side and a lower-point version of the $qg$ basis on the other.
The last two quark channels also factorise onto
lower-point versions of the $qg$ basis on the right,
while producing $qQ$ bases on the left.

\begin{figure}[t]
\begin{align*}
\text{(a)} &~~\scalegraph{0.86}{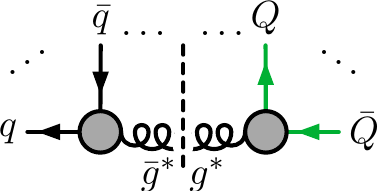} \qquad
\text{(b)}~~\scalegraph{0.86}{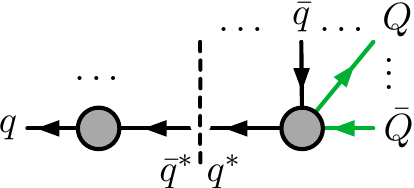} \qquad
\text{(c)}~~\scalegraph{0.86}{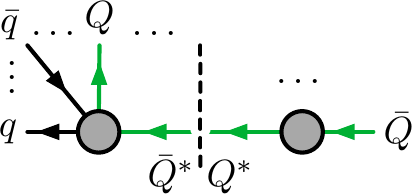} \\ & \qquad~\;\qquad
\text{(d)}~~\scalegraph{0.86}{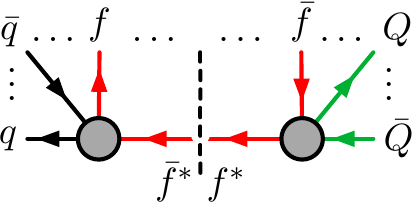} \qquad \quad
\text{(e)}~~\scalegraph{0.86}{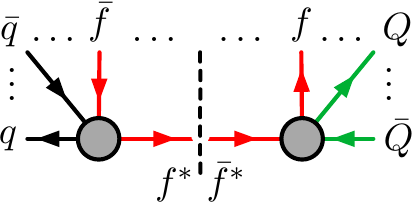}
\end{align*}
\vspace{-15pt}
\caption{\small Factorisation channels of $qQ$-basis colour-ordered amplitude}
\label{fig:qQFactorization}
\end{figure}

Finally, a $qQ$ basis allows five non-zero factorisation limits
of the form $s_{qP} = s_{QR}$, shown in \fig{fig:qQFactorization}:
the particle can be a gluon, a quark of either flavour $q$ or $Q$,
or a quark or antiquark of a third flavour.
Considering when an ordered amplitude $A(q,\pi,\rho,\bar{Q})$
gives a non-vanishing residue in these five cases, we obtain
\begin{subequations} \begin{align}
   {\cal U}_{P,R}\big[{\cal B}_{P \cup R}^{q,Q}\big] &
    = \Big\{ (\pi,\rho) ~\Big|~
             (q)\oplus\pi\,\in \o{\cal Q}(F_P) \shuffle S_{G_P} ,~
             \rho\oplus(\bar{Q})\,\in \o{\cal Q}(F_R) \shuffle S_{G_R}
      \Big\} \nn \\* &
    = {\cal B}_{P}^{q,\bar{g}^*}\!\!\times {\cal B}_{R}^{g^*\!,Q} ,
\label{qQBasisFactorization1} \\
   {\cal U}_{P,R}\big[{\cal B}_{P \cup R}^{q,Q}\big] &
    = \Big\{ (\pi,\rho) ~\Big|~
             \pi \in {\cal Q}(F_P) \shuffle S_{G_P} ,~
      (q^*)\oplus\rho\oplus(\bar{Q})\,\in \o{\cal Q}(F_R) \shuffle S_{G_R}
      \Big\} \nn \\* &
    = {\cal B}_{P}^{q,\bar{q}^*}\!\!\times {\cal B}_{R}^{q^*\!,Q} ,
\label{qQBasisFactorization2} \\
   {\cal U}_{P,R}\big[{\cal B}_{P \cup R}^{q,Q}\big] &
    = {\cal B}_{P}^{q,\bar{Q}^*}\!\!\times {\cal B}_{R}^{Q^*\!,Q} ,
\label{qQBasisFactorization3} \\
   {\cal U}_{P,R}\big[{\cal B}_{P \cup R}^{q,Q}\big] &
    = \Big\{ (\pi,\rho) ~\Big|~
      (q)\oplus\pi\oplus(\bar{f}^*) \in \o{\cal Q}(F_P) \shuffle S_{G_P} ,~
      (f^*)\oplus\rho\oplus(\bar{Q})\,\in \o{\cal Q}(F_R) \shuffle S_{G_R}
      \Big\} \nn \\* &
    = {\cal B}_{P}^{q,\bar{f}^*}\!\!\times {\cal B}_{R}^{f^*\!,Q} ,
\label{qQBasisFactorization4} \\
   {\cal U}_{P,R}\big[{\cal B}_{P \cup R}^{q,Q}\big] &
    = {\cal B}_{P}^{q,f^*}\!\!\times {\cal B}_{R}^{\bar{f}^*\!,Q} .
\label{qQBasisFactorization5}
\end{align} \label{qQBasisFactorization}%
\end{subequations}

\subsubsection{Colour-ordered splitting}
\label{sec:ColourSplittingProof}

Here we show that the bases of colour-ordered amplitudes described in this
section indeed satisfy the ``colour-ordered splitting'' relation, equation
\eqref{eq:BasisSplittingConstraint}. In words this requires us to show that all
elements of an $n$-point basis with particle $i$ before particle $j$ are exactly
those found when taking an $(n-1)$-point basis and performing all possible ordered
splittings producing this pair. As all bases contain each splitting, we consider
each limit in turn.

\paragraph{$qg$ and $gg$ splitting.}

If one of particles $i$ and $j$ is a gluon, then we can easily show that all
bases respect the ordered splitting. Note that the associated $(n-1)$-point basis
is that with one fewer gluon. Importantly, the $n$-point basis can then be thought
of as inserting the remaining gluon in all possible positions in the $(n-1)$-point
basis. This necessarily includes an insertion adjacent to the other particle and so
the colour-ordered splitting relation~\eqref{eq:BasisSplittingConstraint} is
satisfied. Comparing \eqns{FactorizeThenDecompose2}{DecomposeThenFactorize2} we
obtain the colour-factor identity~\eqref{CqgFactorization1} in the $qg$ case and
relation~\eqref{CggFactorization} in the $gg$ case.

\paragraph{$q\bar{q}$ splitting.}
The remaining case is more complicated, as for the $q\bar{q}$ splitting
certain terms in the derivation, specifically in \eqn{DecomposeThenFactorize2},
may be vanishing. This arises due to a property of the allowed quark brackets in
the co-unitary bases. Specifically, all enclosed quark brackets only occur in
one orientation, familiar from the Melia basis.
In this splitting, a $q\bar{q}$ pair is necessarily produced from a gluon.
Given that it respects the orientation criteria, an insertion of an adjacent
quark pair $\{q,\bar{q}\}$ is as legal as a single gluon insertion in all bases.
Indeed, the consistency of a quark bracket structure is never disturbed by either,
so the elements of the two bases are in one-to-one correspondence
as required in \eqn{eq:BasisSplittingConstraint}.

If the splitting gluon is found in an enclosed position, we end up with only one
term when equating \eqns{FactorizeThenDecompose2}{DecomposeThenFactorize2}.
Considering Melia's basis, this holds for the colour factors~\eqref{JOcolor} by
construction. If the splitting corresponds to an unenclosed gluon, then we end
up with both terms. This is then the origin of the $\theta$ in 
the colour-factor identity~\eqref{CqqFactorization2}.

\subsubsection{Linear independence}
\label{sec:basisproof}

Finally, let us now demonstrate the linear independence of the sets
of ordered amplitudes proposed in this work.
This is sufficient to show that these sets form bases,
given that the sets have the right counting $(n-2)!/k!$.
To simplify the argumentation,
we further assume the linear independence of
the Melia basis as established by \rcites{Melia:2013bta,Melia:2013epa}.

First, we note that all three-particle amplitude bases
are trivially independent as they involve only one element.
We wish to work inductively,
assuming that for all $m < n$ our sets of $m$-point amplitudes
form bases of their respective kinematic space.
Precisely, we assume that the only values of $\alpha_\chi$
that solve the equation
\be
    \sum_{\chi \in \mathcal{B}^{a, b}_X} \alpha_\chi
    A\!\left(\scalegraph{0.86}{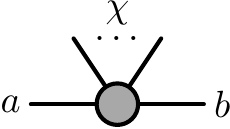}\right) = 0,
\ee
where all amplitudes are $m$-point, are the trivial solutions $\alpha_\chi=0$.
Now we look to find possible solutions $\alpha_\sigma$ of the equation
\be
    \sum_{\sigma \in \mathcal{B}^{1, n}_{P \cup R}}\!\alpha_\sigma
    A\!\left(\scalegraph{0.86}{A1n1permsExposed}\right) = 0 ,
\label{LinearIndependence}
\ee
where we consider the pair of stretched particles~$(1,n)$ to be anything other
than $(q, \bar{q})$. To find solutions, we once again take a factorisation limit
that separates the particles into the two sets $\{1\} \cup P$ and
$R \cup \{n\}$. Importantly, we note that each $\alpha_\sigma$ must turn up at
least in some limit of this form, as every colour-ordered amplitude contains at
least one channel in which it has a non-vanishing residue. ({For this reason we
  rely on the independence of the Melia basis: it contains ordered amplitudes
  that do not have such factorisation channels.}) By co-unitarity of our
amplitude sets, such a limit results in
\be
    \sum_{\pi \in \mathcal{B}^{1, \bar{p}}_P} 
    \sum_{\rho \in \mathcal{B}^{p, n}_R}
    \alpha_{\pi \oplus \rho}
    A\!\left(\scalegraph{0.86}{A1p1permExposed}\right)
    A\!\left(\scalegraph{0.86}{Apn1permExposed}\right) = 0 .
\ee
As our induction hypothesis states that
there are no relations between $A(1,\pi,\bar{p})$,
the coefficient of each ordered amplitude must vanish independently,
so we have
\be
    \sum_{\rho \in \mathcal{B}^{p, n}_R} \alpha_{\pi \oplus \rho}
    A\!\left(\scalegraph{0.86}{Apn1permExposed}\right) = 0
\ee
for each permutation~$\pi$.
However, again by hypothesis,
the only solutions to this equation are $\alpha_{\pi \oplus \rho}=0$.
As this works for all possible limits,
this shows that the coefficients~$\alpha_\sigma$ in \eqn{LinearIndependence}
are indeed all zero, and so the elements are linearly independent. 

In summary, we have constructed amplitude sets with the correct number of elements
that satisfy the factorisation property~\eqref{eq:ColourFactorization}.
This guarantees that the sets are valid bases.

\section{Loop-level applications}
\label{sec:loop}

In \rcite{Ochirov:2016ewn} we presented a general full-colour construction
for loop amplitudes in Yang-Mills theory from generalised unitarity cuts,
which had been used in the two-loop calculation of \rcite{Badger:2015lda}.
In this section we outline the extension of this construction to QCD.

The master formula stays the same as in \rcite{Ochirov:2016ewn}:
\be
   {\cal A}^{(L)}_n = i^{L-1}\!
      \sum_{i\,\in\,\text{KK-indep.\,1PI\,graphs}}
      \int\!\frac{d^{LD} \ell}{(2\pi)^{LD}}
      \frac{C_i\,\Delta_i}{S_i \prod_{l \in i} D_l} .
\label{LoopColour}
\ee
The summation here is over the topologies of the ordered unitarity cuts,
from which the kinematic numerators~$\Delta_i$ are obtained~\cite{Ossola:2006us,
Mastrolia:2011pr,Badger:2012dp,Zhang:2012ce,Mastrolia:2012an,Ita:2015tya}.
The cuts are constructed with tree-level amplitudes as vertices, and for each
such tree only a set of KK-independent orderings need be retained.
The denominators involve the graph symmetry factors~$S_i$
(calculated in the unordered sense)
and the propagator denominators~$D_l$ that were put on shell
to compute the cuts and therefore the numerators~$\Delta_i$.
These numerators may contain poles with respect to the external momenta
but only polynomial dependence on the loop momenta.

The crucial content of the above formula
is the specification that the colour factors~$C_i$
are exactly inherited from the tree-level colour decompositions
inside the unitarity cuts.
This is a simple solution to the a priori puzzling question
of how to combine the physical information from unitarity cuts
into a full-colour integrand without double counting.

In the purely gluonic case the formula~\eqref{LoopColour} means that
one may fix any two edges of every vertex
and sum only over the permutations of the remaining edges,
such that the colour structures inside the vertex
are given by the comb-like colour structures $C_\text{DDM}(g,\sigma,g')$
``stretched'' by the fixed edges. The choice of which two edges to fix is a
priori arbitrary as this will always result in KK independent sets. However, in
\rcite{Ochirov:2016ewn} we show that this freedom can be exploited to further
simplify the loop-amplitude construction with ``stretch'' choices tailored to
specific loop topologies.

In this paper we have formulated a complete set of KK-independent bases and
colour decompositions for QCD tree amplitudes, which now allow a similarly
flexible application of the construction~\eqref{LoopColour} to loop amplitudes in
QCD (or similar gauge theories with matter).
That is, at any stage of a unitarity-based calculation
only the KK-independent cut orderings need be considered,
which correspond to arbitrarily chosen ``stretches''
by two edges of the cut diagram vertices.
The resulting numerators are then dressed with the colour structures that are
sewn from comb-like structures with occasional $C_\text{JO}(q,\sigma,\bar{q})$
elements.

\paragraph{One-loop colour decomposition.}

Let us now illustrate the full-colour approach at one-loop for QCD amplitudes
with external quarks.
Without loss of generality, we restrict to the case of all external quark pairs
having different flavours, as all other cases may be computed by
antisymmetrisation over the distinct-flavour case. For example, an $n$-point
amplitude with four identical quarks can be expressed, irrespectively of the loop
order, as
\be
   {\cal A}\!\left(\scalegraph{0.86}{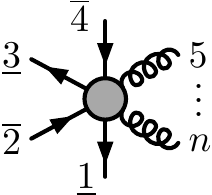}\right)
    = {\cal A}\!\left(\scalegraph{0.86}{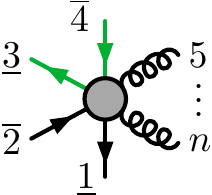}\right)
    - {\cal A}\!\left(\scalegraph{0.86}{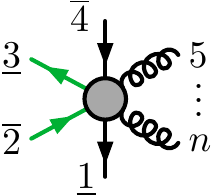}\right) .
\label{FlavorPermutation}
\ee
Here the distinct-flavour amplitudes on the right-hand side are taken with equal
quark masses but with a relative sign, implementing the fermionic antisymmetry
of the like-flavour amplitude on the left-hand side.

\begin{figure}[t]
\centering
\begin{minipage}[c][92pt]{418pt} \hspace{8.2pt}
   \includegraphics[width=391pt,trim={0pt 105pt 0 0},clip=false]{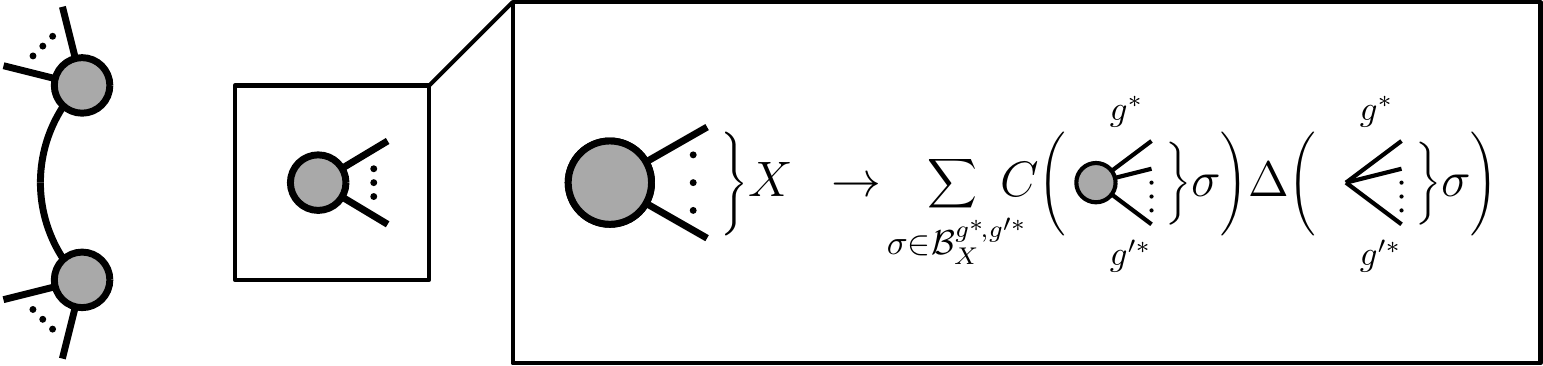} \\ $~~\;\!\!$
   \includegraphics[width=381pt,height=90.7pt,trim={2pt 2pt 30pt 3pt},clip=true]{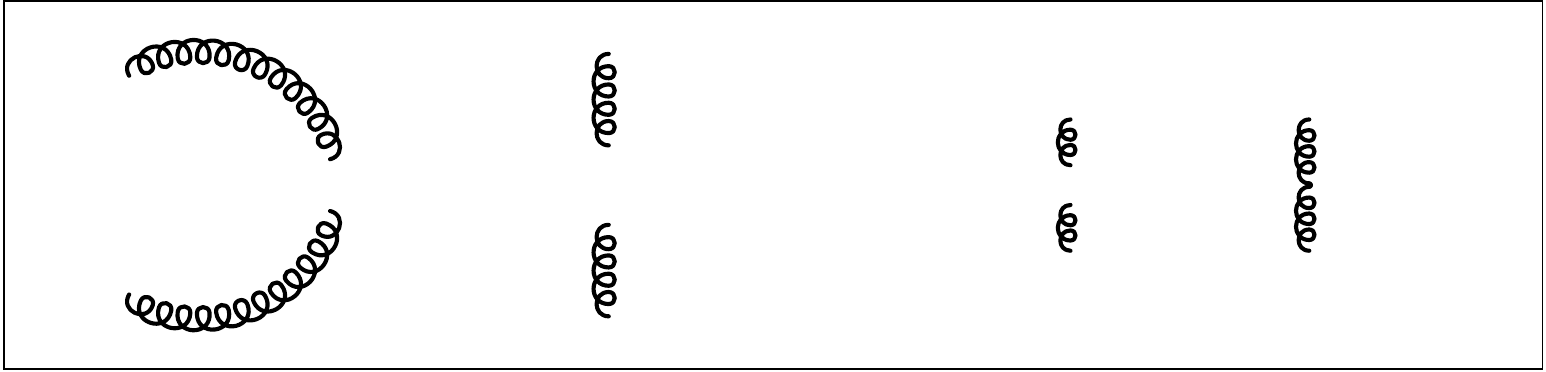}
\end{minipage}
   \caption{\small Inserting a gluon-gluon stretch tree basis into coloured cuts of a one-loop amplitude.}
\label{fig:Color1loop1topo1}
\end{figure}

Working in $D=4-2\epsilon$ dimensions,
the one-particle irreducible topologies that one should consider in \eqn{LoopColour}
at one loop have up to five vertices, each with three or more edges.
Two of these edges are loop-momentum dependent, to which we shall refer
as ``loop edges'' and the rest correspond to the external particles.
For each such topology, one then dresses it in all possible ways with
full-colour tree amplitudes to find the set of possible unordered unitarity cuts.
From these diagrams we compute the symmetry factor $S_i$ in \eqn{LoopColour}.
For each diagram, one can now make a choice of the associated set of
KK-independent ordered unitarity cut diagrams that are summed over
in \eqn{LoopColour} and correspond to the numerators~$\Delta_i$.
In this one-loop case, we choose to stretch the constituent tree amplitudes
across the loop edges, which specifies the KK-independent basis of each corner
to be the ones of this paper.
This choice brings two advantages. First, none of the ordered topologies
have legs pointing inside the loop, and so all numerators
can be readily associated with the leading-colour ordered amplitudes.
Secondly, all kinematic factorisation limits which intertwine the loop-dependent
numerators are also respected by the colour factor through the colour
factorisation relation~\eqref{eq:ColourFactorization}.
Therefore, all colour-ordered numerators related by factorisation
come with the same colour factor.

\begin{figure}[t]
\centering
\begin{minipage}[c][92pt]{418pt} \hspace{8.2pt}
   \includegraphics[width=391pt,trim={0pt 105pt 0 0},clip=false]{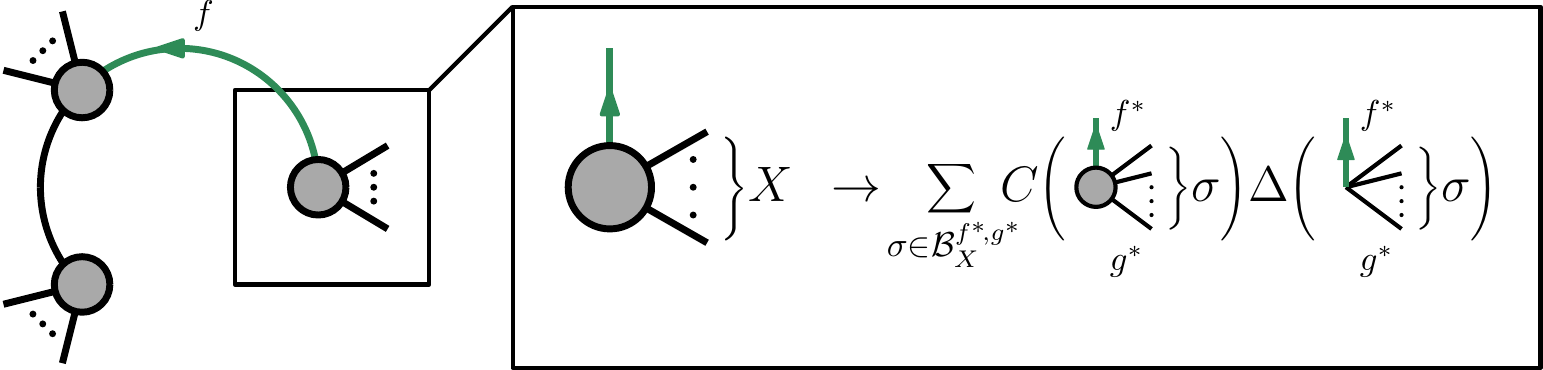} \\ $~~\!$
   \includegraphics[width=383pt,height=93pt,trim={2pt 2pt 30pt 6pt},clip=true]{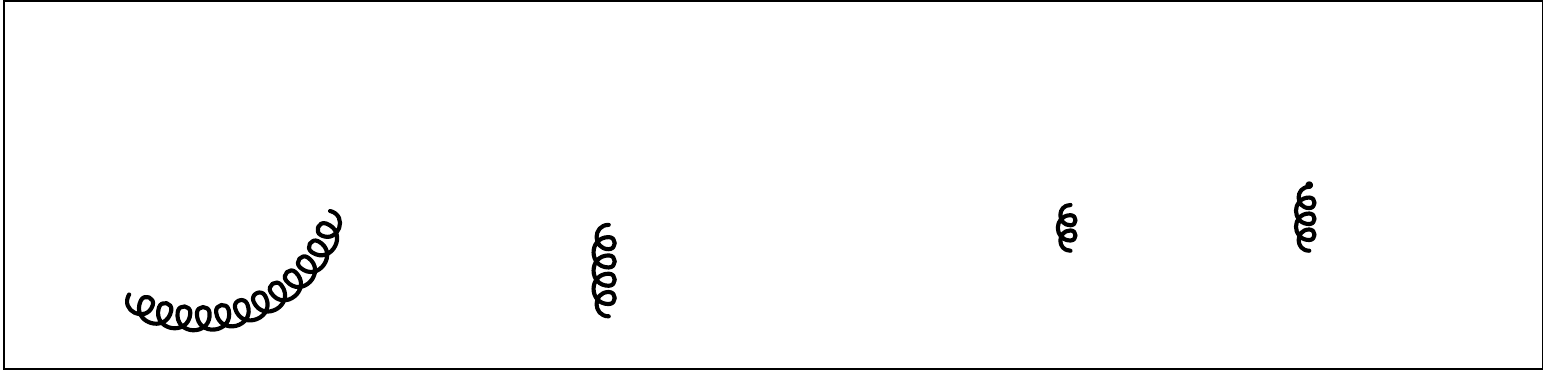}
\end{minipage}
\caption{\small Inserting a quark-gluon stretch tree basis into coloured cuts of a one-loop amplitude.}
\label{fig:Color1loop1topo3}
\end{figure}

\begin{figure}[t]
   \centering
   \includegraphics[width=\textwidth,trim={0pt 5pt 0 0},clip=false]{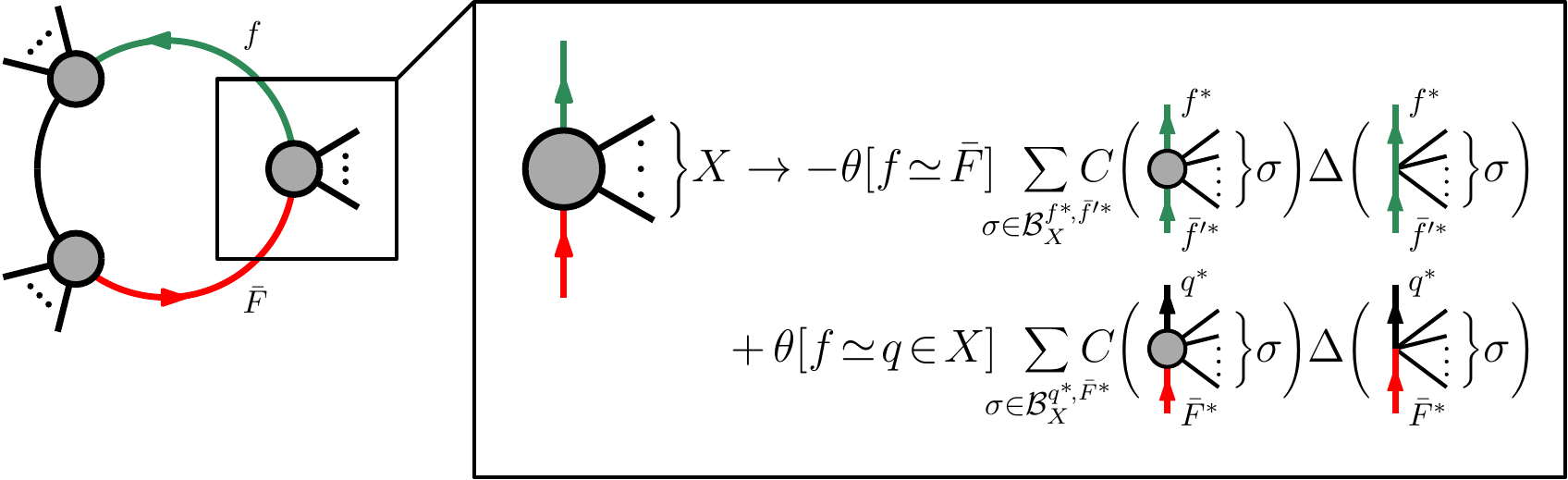}
   \caption{\small Inserting a quark-quark stretch tree basis into coloured cuts
of a one-loop amplitude. The $\theta$ functions are non-zero when the two
flavours are the same.
The first term is present if the stretched quarks are of the same flavour.
The second term is present if the stretched flavour
is also in the set of external flavours.}
\label{fig:Color1loop1topo2}
\end{figure}

Let us now consider how this procedure captures the combinatorics of multi-quark
one-loop amplitudes. In contrast to the purely gluonic case, there are two
novelties. The most evident is that the two ``loop edges'' across which the tree
amplitudes are stretched can now correspond to any two different particles in
the theory. The second is that, due to internal quarks running inside the loop,
the tree amplitudes in the vertices cannot be reduced to distinct flavour
amplitudes.
We shall work through these details by considering each possible type of stretch
in turn. For every vertex inside a cut, its loop edges may correspond to either
\begin{itemize}[leftmargin=\parindent]
\item two internal gluons, as depicted in \fig{fig:Color1loop1topo1};
\item one quark and one gluon, illustrated in \fig{fig:Color1loop1topo3};
\item two internal quarks, as shown in \fig{fig:Color1loop1topo2}
      for the case of the like-flavour edges.
\end{itemize}
In order to concretely discuss the details we discuss these three cases using
the example of two-particle cuts of an $n$-point two-quark amplitude at one
loop. Similar to the adjoint case of~\rcite{Ochirov:2016ewn}, we shall see how the
symmetry factors cancel in the construction. To do this, we first organise the
contributions to \eqn{LoopColour} into colour-dressed numerators
corresponding to an unordered graph;
we denote such numerators by $\tilde{\Delta}_i$.

Consider an $s_{12}$-channel bubble topology,
in which a purely gluonic loop is exposed.
The symmetry factor of this bubble is 2.
The coloured numerator is given by
\beal
   \frac{1}{2} \tilde{\Delta}\!\left(\scalegraph{0.86}{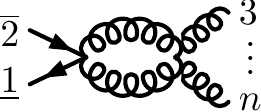}\right)
    = \sum_{\sigma \in S_{n-2}}
      C\!\left(\scalegraph{0.86}{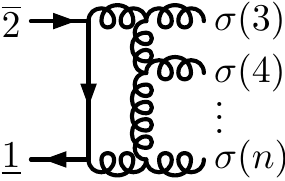}\right)
      \Delta\!\left(\scalegraph{0.86}{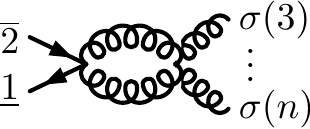}\right) .
\label{AAqqgnDelta1}
\eeal
The set of permutations over which one sums in each corner is given by the bases
with two fixed gluon legs. In the concrete example of \eqn{AAqqgnDelta1},
the explicit permutations and the colour factors on the right-hand side of the
cut are dictated by the DDM decomposition~\eqref{DDM}.
The colour factor on the left-hand side at this point
is also a simple comb-like structure involving both quarks.
The sum over the permutations $\u{1} \leftrightarrow \o{2}$ on the left
naturally produces two copies of each colour-ordered numerator,
which when considered under the integral sign cancel the symmetry factor~2.
This may be familiar from the purely gluonic case~\cite{Ochirov:2016ewn}.
This property relies on the permutation sum generating two copies of each term
related only by a reflection across the axis of the bubble. For this to hold in
more general cases with multiple quark lines on either side, one must take care
to use a basis with a quark bracket signature such that the basis is invariant
under the exchange $[] \leftrightarrow \{\}$.\footnote{In \secs{sec:qQ}{sec:qg},
when we allowed the unnested quark brackets to come in both combinations $\{\}$
and $[]$, we chose the signature for all the nested quark brackets
to remain canonical.
This convention may be switched to have all the nested brackets
follow the signature of their enclosing bracket, \ie
$ \qob{black}\!\dots\!\qo{green}\!\dots\!\qc{green}\!\dots\!\qcb{black} \to
  \qob{black}\!\dots\!\qob{green}\!\dots\!\qcb{green}\!\dots\!\qcb{black} $.
At one loop, this choice allows the colour decompositions
to respect the reflection invariance of the bubbles with internal gluons.
This convention is also consistent with \rcite{Kalin:2017oqr}.
}

Next, consider a bubble cut which involves two distinct gluonic and
fermionic loop lines, whose symmetry factor is unity.
Flavour conservation implies that the flavour of the internal quark line
coincides with an external quark pair split by the unitarity cut.
For such a bubble in the two-quark amplitude,
the colour decomposition presented in \fig{fig:Color1loop1topo3}
should be applied to both sides of the cut.
Schematically, this gives
\beal
   \tilde{\Delta}\!\left(\scalegraph{0.86}{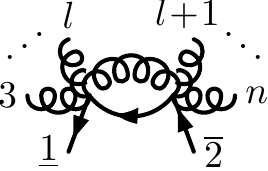}\right)
    = \sum C\!\left(\scalegraph{0.86}{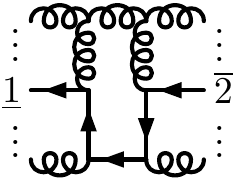}\right)
      \tilde{\Delta}\!\left(\scalegraph{0.86}{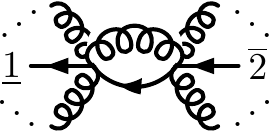}\right) ,
\label{AAqqgnDelta3}
\eeal
where the sum goes over all permutations of the left- and right-hand particle sets.
Due to the absence of differently flavoured quark lines,
the loop colour factors above are sewn from two comb structures,
which are a special case of the $qg$ colour factors~\eqref{qgColor}.

Finally, let us consider the case where both loop legs are fermionic, where we
need to employ the tactics of \fig{fig:Color1loop1topo2}. Here we face the
subtlety where the external and internal fermion lines may indeed be of the same
flavour. For a concrete
example we choose the quark-loop version of the $s_{12}$-channel bubble
topology. We shall label the flavour of the quark loop by $f$ which may or may
not be the same flavour as the external line. In this case we can write the
colour-dressed numerator as
\beal
   \tilde{\Delta}\!\left(\scalegraph{0.86}{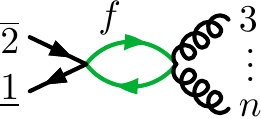}\right)
    = \sum_{\sigma \in S_{n-2}} \Bigg\{
      {-}C\!\left(\scalegraph{0.86}{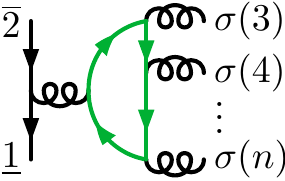}\right)
      \Delta\!\left(\scalegraph{0.86}{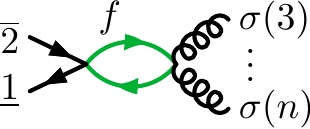}\right) & \\
    + \theta[f \simeq \u{1} \simeq \o{2}]
      C\!\left(\scalegraph{0.86}{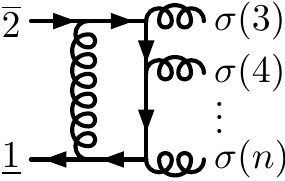}\right)
      \Delta\!\left(\scalegraph{0.86}{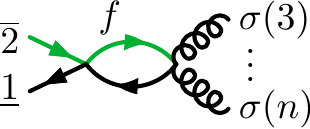}\right) &
      \Bigg\} ,
\label{AAqqgnDelta2}
\eeal
where the theta function encodes the fact that the second term only contributes
if $f$ is the same flavour as particles $\u{1}$ and $\o{2}$.
As the two internal quark lines are distinct, the symmetry factor is unity.
The contributions in the first line of \eqn{AAqqgnDelta2}
enter because the loop edges are two quarks of the same flavour~$f$.
The gluonic permutations on the right-hand side of the bubble 
generate simple DDM comb-like colour factors.
The left-hand side of the bubble does not require any permutations
due to the fixed quark signature of the four-quark Melia basis,
and the corresponding colour factor is familiar from \eqn{A4qqQQ}.
The second line of \eqn{AAqqgnDelta2} only contributes
if the internal quarks are of the same flavour as the external ones,
in accord with the construction~\eqref{FlavorPermutation}
of the like-flavour amplitude from the distinct-flavour ones.
The fixed quark signature of $qQ$-stretch basis on the left-hand side of the bubble
still requires no permutations,
apart from the gluonic ones from the right-hand side.
The colour factors are comb-like on both sides
due to the absence of additional quark lines.

Of particular interest is the relative fermionic sign
between the two lines of \eqn{AAqqgnDelta2}.
The first line corresponds to the so-called $N_f$ contributions
to the amplitude due to a closed quark loop,
whereas in the second line the quark loop is not closed.
Indeed, the associated unitarity cut contains a loop-momentum dependent
channel that corresponds to a gluonic line. Due to colour factorisation,
this is already manifest in the colour factor.
We observe that the resulting (relative) fermionic sign
follows through from the flavour-permutation construction~\eqref{FlavorPermutation}.
It is a general feature of our approach
that the fermionic signs from flavour permutations feed into unitarity cuts
and produce consistent signs for topologies with fermion loops.

In the above examples, we have reduced the colour factors
to explicit diagrams with only three-point vertices.
However, their more general feature is that
they are all given by a diagram which is a ring
built from the $C_\text{DDM}(g,\sigma,g')$ and $C_\text{JO}(q,\sigma,\bar{q})$ building blocks.
We emphasise that this is a general feature of the one-loop decomposition,
enabled by the colour factorisation relation~\eqref{eq:ColourFactorization}
of the tree-level decompositions.
As the constituent tree amplitudes in each cut are stretched across the loop legs,
and the tree colour factors themselves factorise into such building blocks,
this construction is naturally inherited by the loop-level factors.

Finally, we note that the presented approach is completely consistent with
the known methods for colour decompositions of multi-quark amplitudes.
For example, the colour-ordered numerators that we have constructed here
can be associated with the left/right-turner and $N_f$ families
of one-loop primitive amplitudes defined in \rcites{Bern:1994fz,Ita:2011ar}.
Furthermore, our results are also completely consistent with K\"alin's recent
one-loop decomposition into such primitive amplitudes~\cite{Kalin:2017oqr},
as can be easily verified by comparing the unitarity cuts. However, the
tree-level results cannot be directly derived from K\"alin's 
decomposition, as it does not involve symmetry factors and so its cuts
correspond to products of tree-level amplitudes with certain terms identified.

\section{Summary and outlook}
\label{sec:outro}

In this paper we have considered the colour structure of tree and one-loop QCD
amplitudes involving any number of distinctly flavoured quark-antiquark pairs.
At tree level, we have derived new bases of ordered amplitudes
that are independent under the Kleiss-Kuijf relations~\cite{Kleiss:1988ne}
and found decompositions of an $n$-point colour-dressed amplitude into these bases.
In combination with the previously known colour decompositions
of \rcites{DelDuca:1999rs} and~\cite{Johansson:2015oia,Melia:2015ika},
our results permit flexible amplitude implementations in terms of
ordered amplitudes with an arbitrary pair of partons fixed next to each other.
At (multi-)~loop level, this flexibility significantly enhances the application
of the loop-colour approach of~\rcite{Ochirov:2016ewn},
as demonstrated by our more detailed exposition at one loop.

Another important aspect of this paper is the recursive unitarity-based approach
that we employ to construct new colour decompositions. It relies on the physical
factorisation properties of the colour-dressed and colour-ordered amplitudes,
which impose certain factorisation relations on the colour factors.
Due to its physical transparency,
this method for handling colour factors
could arguably be considered as advantageous with
respect to closed-form expressions.
For instance, the implementation strategy where colour factors are recursed down to
fundamental three-point vertices has already proven its simplicity in the recent
two-loop computations at leading colour~\cite{Abreu:2018jgq,Abreu:2019odu} by
one of the authors.

Apart from the proposed utility of our results for
future multi-loop QCD calculations via on-shell methods,
there are a number of other natural applications.
In particular, the colour-kinematics duality~\cite{Bern:2008qj,Bern:2010ue}
present in QCD~\cite{Johansson:2014zca,Johansson:2015oia}
can be used to uplift the presented colour decompositions
to new representations of gravitational amplitudes
in QCD minimally coupled to general relativity~\cite{Plefka:2018zwm}.
Moreover, the duality implies the kinematic-dependent amplitude
relations~\cite{Bern:2008qj,Johansson:2015oia,delaCruz:2015dpa},
which could be used to further reduce the bases formulated here. It could
be interesting to explore whether such reduced bases could provide additional
flexibility in the application of such amplitude relations to loop
amplitudes~\cite{Badger:2015lda,Primo:2016omk,Ochirov:2017jby}.

Our loop-colour approach~\cite{Ochirov:2016ewn} is a tool
for constructing the loop integrand in gauge theory.
The colour factorisation perspective taken in this paper
naturally organises loop integrand contributions in this approach, as we have
explicitly demonstrated at one loop.
However, there are a number of available analytical full-colour results
for two-loop amplitudes in pure Yang-Mills theory~\cite{Badger:2019djh},
its ${\cal N}=4$ supersymmetric extension~\cite{Carrasco:2011mn,Abreu:2018aqd,Chicherin:2018yne}
and ${\cal N}=2$ supersymmetric QCD~\cite{Duhr:2019ywc},
which organise the integrated amplitudes
in terms of the more traditional trace basis,
subject to known colour redundancies~\cite{Edison:2011ta}.
A natural question is then how factorisation could guide the organisation of
colour structures, not only of the integrand of loop amplitudes, but also after integration.

\begin{acknowledgments}

We thank Samuel Abreu, Harald Ita and Henrik Johansson for helpful
conversations. We also thank Fernando Febres Cordero, Henrik Johansson and
Gregor K\"alin for comments on the manuscript. AO's research is funded by the
European Union's Horizon 2020 research and innovation programme under the Marie
Sk{\l}odowska-Curie grant agreement 746138. BP is grateful for the support from
the Pauli Center for Theoretical Physics. The work of B.P. is supported by the
French Agence Nationale pour la Recherche, under grant ANR–17–CE31–0001–01.

\end{acknowledgments}

\appendix
\section{Colour Feynman rules and colour ordering}
\label{app:feynmanrules}

Here we recall the standard colour vertices
defined with respect to planar ordering:
\be
   \scalegraph{1.0}{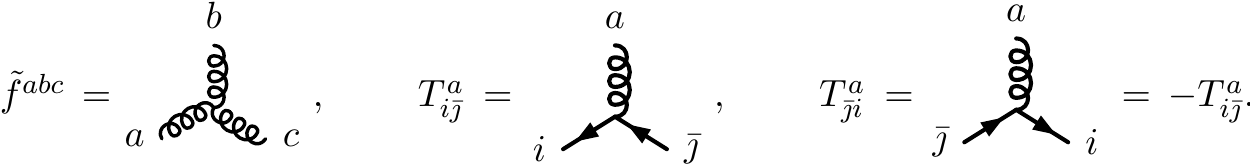}
\ee
A convenient way to normalise the gauge-group generators is by imposing
\be
   \Tr(T^{a} T^{b})=\delta^{ab}, \qquad [T^a,T^b]=\tf^{abc}T^c
   \qquad \Rightarrow \qquad
   \tf^{abc} = \tr\!\big([T^a,T^b]T^c\big) .
\ee

The purely kinematic ordered Feynman rules may be found \eg in \rcite{Dixon:1996wi}
but their specific form is irrelevant for the purely colour-algebraic results
of the present paper.
The general applicability of the concept of colour ordering relies on the fact
that in a unitary gauge theory a matter particle in a complex group representation
can be projected to two real particles in the adjoint representations
with the same kinematic Feynman diagrams but with all generators replaced
by the structure constants.
Colour ordering can then be performed in the same way as for gluons,
thereby defining the ordered amplitudes $A(\sigma(1),\dots,\sigma(n))$ with matter
as gauge-invariant kinematic coefficients of the fundamental traces
$\tr(T^{a_{\sigma(1)}} \cdots T^{a_{\sigma(n)}})$.
Once ordered in this way in the adjoint representation, these amplitudes can be
dressed with the arbitrary-representation colour coefficients,
on which we concentrate in the bulk of the paper.

\section{Colour-unitarity checks}
\label{sec:qqunitarity}

Here we show that the colour factors given by \eqn{JOcolor} obey
the leg-exchange relations~\eqref{CqqFactorization2},
\eqref{CqgFactorization} and~\eqref{CggFactorization}.
The former is satisfied by construction
as illustrated by the first line of \eqn{A5qqColor}:
when a quark pair is not separated by any other particles,
its colour factor factorises explicitly onto the adjoint index of
the intermediate gluon via
$\qo{black}q|T^{a_{g*}}|\bar{q}\qc{black} = T^{a_{g*}}_{i_q \bar\imath_{\bar{q}}}$.

The gluonic colour-factor relation~\eqref{CggFactorization}
is also obeyed automatically due to the commutation relation
between the tensor-representation generators,
\be
   [\Xi^a_l,\Xi^b_l] = \tf^{abc} \Xi^c_l .
\label{XiCommutation}
\ee
Indeed, when two gluons are next to each other, the corresponding colour factor
necessarily involves two $\Xi$'s of the same nestedness level $l$,
so the commutation relation generates the adjoint vertex
to the intermediate gluon, as in \eqn{CggFactorization}.

Now let us consider
the quark-gluonic exchange relation~\eqref{CqgFactorization} in detail.
Drawing only the relevant parts of the colour diagrams for the $qg$ case,
we compute in full generality
\begin{align}
 & C\!\left(\scalegraph{0.8}{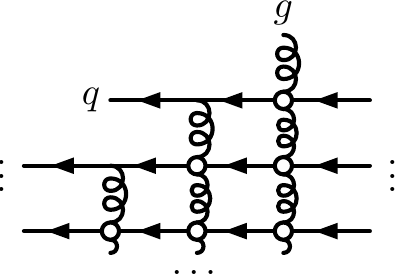}\right)
    - C\!\left(\scalegraph{0.8}{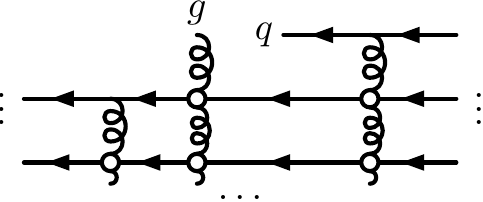}\right) \\ &
    = C\!\left(\scalegraph{0.8}{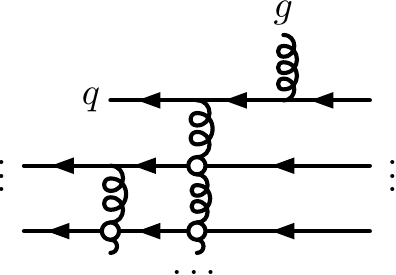}\right)
    + C\!\left(\scalegraph{0.8}{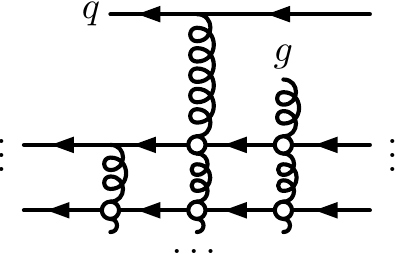}\right)
    - C\!\left(\scalegraph{0.8}{Cgq}\right) \nn \\ &
    = C\!\left(\scalegraph{0.8}{CqgTop}\right)
    + C\!\left(\scalegraph{0.8}{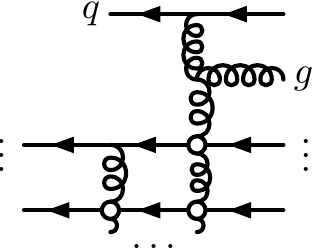}\right)
    = C\!\left(\scalegraph{0.8}{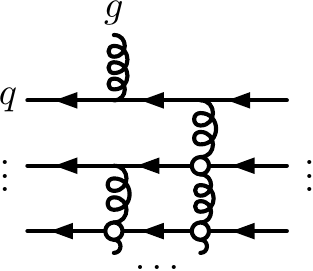}\right) . \nn
\end{align}
Here we first used the additive nature of the tensor-representation generators
that is evident from the definition, namely $\Xi^a_l = T^a_l + \Xi^a_{l-1}$,
and then applied
the commutation relations~\eqref{XiCommutation} and~\eqref{Commutation}.
In the resulting diagram the original quark~$q$ and gluon~$g$
are attached to a lower-point colour factor of the type~\eqref{JOcolor}
in exactly the same way as required
by the leg-exchange identity~\eqref{CqgFactorization1}.
Finally, the corresponding $g\bar{q}$ condition~\eqref{CqgFactorization2}
follows exclusively from the aforementioned recurrence relation
$\Xi^a_l = T^a_l + \Xi^a_{l-1}$:
\be
   C\!\left(\scalegraph{0.8}{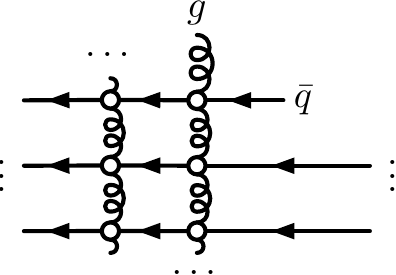}\right)
    - C\!\left(\scalegraph{0.8}{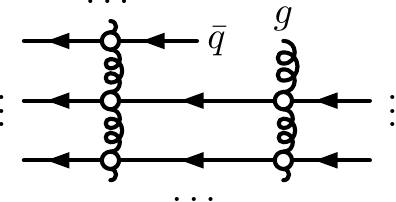}\right)
    = C\!\left(\scalegraph{0.8}{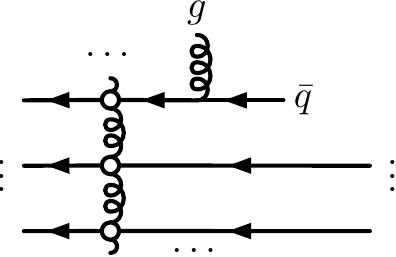}\right) .
\ee
In other words, the left-right asymmetric choice
of the colour diagrams in \eqn{JOcolor}
makes the $g\bar{q}$-exchange property~\eqref{CqgFactorization2} manifest
for the price making the $qg$-exchange property~\eqref{CqgFactorization1} hidden
behind a couple of commutation relations.

\bibliographystyle{JHEP}
\bibliography{references}

\end{document}